\documentclass[aps,prd,superscriptaddress,nofootinbib,eqsecnum,twocolumn]{revtex4-1}

\pdfoutput=1

\usepackage{amsfonts}
\usepackage{amsmath}
\usepackage{amssymb}
\usepackage{graphicx,color}
\usepackage{float}
\usepackage{hyperref}
\usepackage{subfigure}
\usepackage{dcolumn}
\usepackage{soul}
\usepackage{ulem}
\usepackage{graphicx}
\usepackage{color}

\begin{document}

\title{Finite size effects in the phase transition patterns of coupled scalar field systems}

\author{Lucas G. C\^amara} \email{prof.lucasgondim@gmail.com}
\affiliation{Departamento de Fisica Teorica, Universidade do Estado do
  Rio de Janeiro, 20550-013 Rio de Janeiro, RJ, Brazil }

\author{Rudnei O. Ramos} \email{rudnei@uerj.br}
\affiliation{Departamento de Fisica Teorica, Universidade do Estado do
  Rio de Janeiro, 20550-013 Rio de Janeiro, RJ, Brazil }
\affiliation{Physics Department, McGill University, Montreal, Quebec, H3A 2T8, Canada}

\begin{abstract}

It is considered in this work the phase transition patterns for a
coupled two-scalar field system model under the combined effects of
finite sizes and temperature.  The scalar fields are taken as
propagating in a $D=4$ Euclidean space with the usual periodic
compactification in the Euclidean time direction (with dimension given
by the inverse of the temperature) and also under a compact dimension
in the space direction, which is restricted to size $L$.  In the latter
case, a Dirichlet boundary condition is considered.  
{}Finite-size variation of the critical temperature for the cases
of symmetry restoration and inverse symmetry breaking are studied. 
At fixed finite-temperature values, the variation of the inverse correlation
lengths with the size $L$ might display a behavior analogous to reentrant 
phase transitions.
Possible applications of our results to physical systems of
interest are also discussed.

\end{abstract}

\maketitle

\section{Introduction} 
\label{intro}

Studies of phase transitions under different conditions, like
temperature, external fields and chemical potential, just to cite a
few examples, have always been of relevance. Modeling and describing
these phase transitions through effective field theories with, e.g.,
scalar fields, find applications in a variety of physical systems of
interest, ranging from condensed matter to cosmology.  Studying the
properties of quantum field theory models involving multiple scalar
fields and understanding how symmetries change in these models when
undergoing phase transitions have gained relevance recently. The
reason for this interest is that these type of models can have
connections ranging from statistical physics and  condensed
matter~\cite{Kosterlitz:1976zza,Calabrese:2002bm,Eichhorn:2013zza,Demler:2004zz,Aharony:2022blf}
to high-energy
physics~\cite{Meade:2018saz,Baldes:2018nel,Matsedonskyi:2020mlz,Matsedonskyi:2020kuy,Bajc:2020yvd,Chaudhuri:2020xxb,Chaudhuri:2021dsq,Niemi:2021qvp,Ramazanov:2021eya,Croon:2020cgk,Schicho:2021gca}.

A model that has been of particular interest is a multiple scalar
field system with a $O(N_\phi) \times O(N_\chi)$, for a Lagrangian
density containing scalar fields $\phi \equiv (\phi_1, \ldots,
\phi_{N_\phi})$ and $\chi \equiv (\chi_1, \ldots, \chi_{N_\chi})$,
with $N_\phi$ and $N_\chi$ components, respectively. This type of
model has long been of interest and have also been studied before in
different
contexts~\cite{Mohapatra:1979qt,Klimenko:1988mb,Bimonte:1995xs,AmelinoCamelia:1996hw,Orloff:1996yn,Roos:1995vm,Pietroni:1996zj,Jansen:1998rj,Bimonte:1999tw,Pinto:1999pg,Pinto:2005ey,Pinto:2006cb,Farias:2021ult}.
Those studies mostly focused on how symmetries in these type of
systems change with the temperature.  The interest on these type of
models derives from the fact that they can display a rich phase
structure depending on the parameter space available for them. In
particular, it is known since the work done by Weinberg in
Ref.~\cite{Weinberg:1974hy} that nontrivial phase transition patterns
can be displayed by these type of models. These phase transitions are,
for example, related to  {\it inverse symmetry breaking} (ISB), i.e.,
the breaking of symmetries at high temperatures,  as well as {\it
  symmetry nonrestoration} (SNR), namely, the persistence of symmetry
breaking at high temperatures.  In this work, we are interested in
investigating the patterns of phase transitions in the above multiple
scalar field model, but including the effects of a finite boundary
along a space dimension in addition to the known effects of
temperature.

Studies of finite-size effects in quantum field theory have long been
of importance, like, for example, in understanding  the questions
related to vacuum energy, e.g., in the Casimir energy in different
topologies~\cite{Ambjorn:1981xw,Blau:1988kv,Cognola:1991qu,Elizalde:1994gf,Elizalde:1995hck}. 
{}Finite-size effects have also been shown to lead to phase transitions (see,
e.g., Ref.~\cite{Khanna:2014qqa} and references there in). This can
happen since space compactifications work similarly to the
introduction of temperature in the Matsubara formalism of finite-temperature 
quantum field theory in Euclidean spacetime and where the
Euclidean time direction is compactified to a finite dimension given
by the inverse of the temperature~\cite{Bellac:2011kqa}. Space
compactifications can then modify the effective masses of the fields
in quantum field theory and affect the symmetry properties of these
fields~\cite{Toms:1979ij}. 

In a practical context, like in most experiments under laboratory
conditions, the limitation of the system size can produce important
boundary effects. The thermodynamic limit in these cases might not
give a reliable result and, in fact alter many critical behaviors of
the system~\cite{Fisher:1972zza,Brezin:1985xx}. Likewise, in the
studies of heavy ion collisions performed, e.g., at the Relativistic
Heavy Ion Collider (RHIC) and at the Large Hadron Collider (LHC), it
has been indicated that the mean-free path of quarks and gluons formed
is not much smaller than the typical fireball
radius~\cite{Romatschke:2016hle}. This indicates that the
thermodynamics of the quark-gluon plasma (QGP) can have sizable
effects from the boundaries of the system and which has motivated many
studies of finite-size effects towards  understanding their
contributions~\cite{Palhares:2009tf,Braun:2011iz,Bhattacharyya:2012rp,Samanta:2017ohm,Mogliacci:2018oea,Kitazawa:2019otp}.

In this paper, we will be interested in studying how the introduction
of a boundary affects the thermodynamics of a coupled two scalar field
system. In particular, we want to focus on the possible emergence of
reentrant phases and symmetry persistence in systems of this
type. This study complements the many previous studies on similar
systems, which, however, up to our knowledge, have not explored how a
boundary might eventually affect the phase transition in this
context. Even though we do not focus on a particular application, this
study might be of relevance in understanding condensed matter systems
that can be well modeled by these type of models in an effective
description, besides, as already cited earlier, of also being of
theoretical interest in general.  In the present study, we make use of
the  nonperturbative resummation of the one-loop order terms, i.e., in
the ring (bubble) resummed
approximation~\cite{Espinosa:1992gq,Amelino-Camelia:1992qfe,Drummond:1997cw,Kraemmer:2003gd,Ayala:2004dx}.
Similar techniques were also previously used to study ISB and SNR, but
in the absence of space
boundaries~\cite{Roos:1995vm,Pietroni:1996zj,Pinto:2005ey,Pinto:2006cb}.
The results obtained here are also compared in the context of the
large-$N$ approximation for the model.

{}For definiteness, we will also consider the case of a Dirichlet
boundary condition, which is motivated by both condensed matter type
of systems where the wave function vanishes at the surface of the
material and does not propagate beyond it. This type of boundary
condition has also been claimed to be the appropriate one to consider
for finite-size effects on the thermodynamics of the
QGP~\cite{Mogliacci:2018oea}. As an additional advantage of using a
Dirichlet boundary condition  is that it allows to obtain simple
analytical approximate equations, which facilitate the analysis and
interpretation of the results. As we are going to show, besides of the
ISB and SNR phenomena, behavior analogous to 
reentrant symmetry breaking, with double
transition points, can also manifest when space dimensions are
constrained to finite sizes.

The remainder of this paper is organized as follows. In
Sec.~\ref{section2}, we review the basics of ISB/SNR for a
$O(N_{\phi})\times O(N_{\chi})$ invariant relativistic scalar field
model in the context of perturbation theory (PT) in the high-temperature
approximation. In Sec.~\ref{section3}, the self-energy corrections to
the fields and that are dependent on the temperature and finite size
are derived. The effective masses entering in our calculations are
then computed. In Sec.~\ref{section4} the renormalized parameters of
the model are discussed and the bubble (ring) resummed masses are
given along also the tadpole equations that give the expectation
values for the fields. Our results are discussed in Sec.~\ref{results}
and the possibility of behavior analogous to 
reentrant phase transitions at high temperatures and under the effects of the boundary
are studied. In Sec.~\ref{largeN} the large-$N$ approximation is
implemented in the model and the results are again compared by varying
the number of components for the fields. {}Finally, in
Sec.~\ref{conclusions} our conclusions are given and possible
applications of our results are discussed. 

\section{The perturbative description of ISB and SNR at finite temperature}
\label{section2}

The prototype model we consider is that of a  $O(N_{\phi})\times
O(N_{\chi})$  invariant relativistic scalar field model, with  $\phi$
and $\chi$ consisting of scalar fields with $N_\phi$ and $N_\chi$
components, respectively.  The interactions are given by the standard
self-interactions among each field species, with  quartic couplings
$\lambda_\phi$ and $\lambda_\chi$, respectively and by a quadratic
(inter) cross-coupling  $\lambda$ between $\phi$ and $\chi$ The
Lagrangian density is given by
\begin{eqnarray}
{\cal L} &=& \frac{1}{2} \left(\partial_\mu \phi \right)^2 -
\frac{1}{2} m_\phi^2 \phi^2 - \frac{\lambda_\phi}{4 !} \phi^4
\nonumber \\ &+& \frac{1}{2} \left(\partial_\mu \chi \right)^2 -
\frac{1}{2} m_\chi^2 \chi^2 - \frac{\lambda_\chi}{4 !} \chi^4 -
\frac{\lambda}{4} \phi^2 \chi^2.
\label{Lphich}
\end{eqnarray}
The potential is always bounded for positive couplings, but the
overall boundness of the potential is still maintained even when the
intercoupling $\lambda$ is negative, provided that
\begin{equation}
\lambda_{\phi} \lambda_{\chi} > 9 \lambda^2, \;\; \lambda_{\phi}
>0,\;\; \lambda_{\chi} > 0,
\label{boundness}
\end{equation} 
and in this case ISB and SNR can emerge at finite temperature, as
first shown in the seminal work in Ref.~\cite{Weinberg:1974hy}. {}For
instance, restricting to the  one-loop approximation and in the high-temperature 
approximation ($m_\phi/T, m_\chi/T \ll 1$),  the thermal
masses for the $\phi$ and $\chi$ fields are simply 
\begin{equation}
M_{i}^2(T) \simeq m^2_i + \frac{T^2}{12}  \left(\frac{N_{i}+2}{6} \,
\lambda_{i} +  \frac{N_{j}}{2} \lambda \right), \;\;{i,j=\phi,\chi}.
\label{MiT}
\end{equation}
Equation~(\ref{MiT}) shows that ISB/SNR can emerge for $\lambda < 0$,
if, for example, $m_{i,j}^2>0$, i.e., we have a symmetric theory at
$T=0$ in both $\phi$ and $\chi$ directions initially and  ISB can take
place in the direction of one of the fields if
\begin{equation}
|\lambda| > \frac{\lambda_i}{N_j}\left (\frac{N_i+2}{3} \right ),
\label{isbcondition}
\end{equation}
since the $T^2$ coefficient in Eq.~(\ref{MiT}) becomes negative and
then, a symmetry breaking like $O(N_i)\times O(N_j) \rightarrow
O(N_i-1)\times O(N_j)$ can occur. Note that the boundness condition
Eq.~(\ref{boundness}) prevents that ISB might come to happen in the
other field direction.  On the other hand, starting with a theory in
the broken phase in both field directions, $m_{i,j}^2 < 0$, under the
condition Eq.~(\ref{isbcondition}), we now have SNR, since there is in
principle no critical temperature for symmetry restoration (SR) in the
$i$-field direction, while the other field suffers SR as usual at a
critical temperature
\begin{equation}
T_{j,c}^2 = -\frac{72  m_j^2}{(N_j+2)\lambda_j + 3 N_i \lambda}.
\label{TcSR}
\end{equation}
In this case, the symmetry changing scheme is $O(N_i-1)\times O(N_j-1)
\rightarrow O(N_i-1)\times O(N_j)$.

A natural question to ask is whether these nonusual symmetry patterns
at high temperature would not be just artifacts of the naive one-loop
approximation and high-temperature approximation.  In principle, it
could well be the case that higher-order terms could restore the usual
SR patterns expected commonly. However, many
previous papers using a variety of nonperturbative methods give
support for
ISB/SNR~\cite{Bimonte:1995xs,AmelinoCamelia:1996hw,Orloff:1996yn,Roos:1995vm,Pietroni:1996zj,Jansen:1998rj,Bimonte:1999tw,Pinto:1999pg,Pinto:2005ey,Pinto:2006cb}
and also in more recent works, like e.g., in
Refs.~\cite{Chaudhuri:2021dsq,Chai:2021tpt,Liendo:2022bmv}. In the
next sections we will explore this problem when in addition to
temperature, finite-size effects are also included.

\section{Effective masses dependence on $T$ and $L$}
\label{section3}

We want to extend the above analysis when now there is a
compactification in one of the space dimensions. In other words, we
want to study the above picture of ISB/SNR when in the presence of
finite-size effects.  {}Following the quantum field theory formalism
in toroidal topologies~\cite{Khanna:2014qqa}, the finite-size effects
can be included by defining the space in a topology
$\Gamma^d_{D}=(\bf{S}^1)^d\times \bf{R}^{D-d}$, where $D$ is the
space-time dimension and $d$ is the number of compactified dimensions,
such that $d\leq D$. Let us see how this can be generalized to the
present problem. {}For this, let us return to a moment to the one-loop
perturbative Eq.~(\ref{MiT}) and express it in terms of the original
momentum integrals. In Euclidean D-dimensional momentum space, we then
have that
\begin{eqnarray}
M_i^2 &=& m_i^2 + \frac {\lambda_i}{2} \left ( \frac {N_i+2}{3} \right
) \int \frac{d^D p}{(2\pi)^D} \frac {1}{p_E^2+ m_i^2} \nonumber
\\ &+&\frac {\lambda}{2} N_j \int \frac{d^D p}{(2\pi)^D} \frac
   {1}{p^2_E+m_j^2},
\label{S1a}
\end{eqnarray}
where $i,\,j$  represent either $\phi$ or $\chi$. {}For each
compactified dimension, with finite lengths $L_a$, $a \leq d$, the
corresponding momentum $p_a$ in that direction is replaced in terms of
discrete frequencies $\omega_{n_a}$, $p_a \to \omega_{n_a}$.  The
discrete frequencies $\omega_{n_a}$ depend on the boundary condition
(BC). Some well know BCs used in the literature under different
contexts are, for example, the periodic, antiperiodic, Neumann and
Dirichlet boundary conditions.  {}For a periodic BC we have 
\begin{equation}
\omega_{n_a} = \frac{2 \pi n_a}{L_a}, \;\;\; n_a \in \mathbb{Z},
\label{PBC}
\end{equation}
for an antiperiodic BC, 
\begin{equation}
\omega_{n_a} = \frac{(2 n_a+1)\pi }{L_a}, \;\;\; n_a \in \mathbb{Z},
\label{APBC}
\end{equation}
for a Neumann BC,
\begin{equation}
\omega_{n_a} = \frac{\pi n_a}{L_a}, \;\;\; n_a \in \mathbb{N},
\label{NBC}
\end{equation}
while for a Dirichlet BC,
\begin{equation}
\omega_{n_a} = \frac{\pi n_a}{L_a}, \;\;\; n_a \in \mathbb{N}_{>0}  .
\label{DBC}
\end{equation}

The periodic and antiperiodic BCs are well known in the context of
quantum field theory at finite temperature, where the (Euclidean) time
gets compactified with size $L\to 1/T$, where $T$ is the temperature
and the corresponding frequencies are given   in terms of the
Matsubara's frequencies for bosons (periodic BC) or fermions
(antiperiodic BC).  Neumann BC applies when, e.g., the derivative of
the wave function vanishes at the boundaries, while for Dirichlet BC
the wave function (or field) is identically zero at the boundaries and
outside the bulk.  A Dirichlet boundary condition (DBC) can be seen
then like an impenetrable barrier.

Thus, for each compactified space dimension, with finite-lengths
$L_a$, we can write the momentum integral in the corresponding
direction as 
\begin{equation}\label{eq15}
\int \frac{dp_a}{2\pi}\rightarrow \frac{1}{L_a}\sum_{n_a},
\end{equation}
and for $d<D$ compactifications, the momentum integrals, which we will
be interested in this work, they are all of the form 
\begin{eqnarray}
I^{(\alpha)}_{D-d}&=&\frac{1}{L_1\cdots L_d} \sum_{n_1, \ldots, n_d}
\nonumber \\ &\times& \int \frac{d^{D-d} p}{(2\pi)^{D-d}} \frac
          {1}{\left(\omega_{n_1}^2+ \cdots + \omega_{n_d}^2+
            p^2_{D-d}+ m_i^2\right)^\alpha}, \nonumber \\
\label{IDd}
\end{eqnarray}
where $p^2_{D-d}$ is the Euclidean momentum in $(D-d)$-dimensions and
$m_i\equiv m_{\phi(\chi)}$.  The momentum integrals in Eq.~(\ref{S1a})
are in particular obtained by setting $\alpha=1$ in Eq.~(\ref{IDd}).
The momentum integral in the remaining $(D-d)$-dimensions in
Eq.~(\ref{IDd}) can be performed in dimensional regularization.
Working in dimension $\delta \equiv D-d$, with $D=4-2\epsilon$ in the
$\overline{\rm MS}$-dimensional regularization scheme, we have for
Eq.~(\ref{IDd}) the result
\begin{eqnarray}
I^{(\alpha)}_{D-d}&=&\frac{1}{L_1\cdots L_d} \left(\frac{e^{\gamma_E}
  \mu^2}{4 \pi}\right)^{\epsilon} \frac{\Gamma(\nu)}{(4 \pi)^{(D-d)/2}
  \Gamma(\alpha) } \nonumber \\ &\times& \sum_{n_1, \ldots, n_d}
\left(\omega_{n_1}^2+ \cdots + \omega_{n_d}^2+ m_i^2 \right)^{-\nu},
\label{IDd2}
\end{eqnarray}
where $\nu=\alpha -(D-d)/2$, $\mu$ is an arbitrary mass regularization
scale (in the  $\overline{\rm MS}$-scheme) and $\gamma_E$ is the
Euler-Mascheroni constant.

Performing the sum on the right hand side of Eq.~(\ref{IDd2}) is
arduous in general. However, the job can get simplified by expressing
those type of sums in terms of an Epstein-Hurwitz zeta
function~\cite{Elizalde:1995hck}
\begin{equation}
G^{c^2}_{d}(\nu;a_1,\ldots,a_d)=\!\!\!\!\sum^{\infty}_{n_1,\ldots,n_d=-\infty}\left(c^2+a_1^2
n_1^2 + \ldots + a_d^2 n_d^2\right)^{-\nu},
\label{Gc}
\end{equation}
where, for example, we can identify $c = m_i$, $a_j$ ($j=1,\ldots d$)
are coefficients that depend on the  BCs (see, e.g.,
Eqs.~(\ref{PBC})-(\ref{DBC})) and $\nu=\alpha-\delta/2$.
{}Following, e.g., Ref.~\cite{Malbouisson:2002am}, we can also express
Eq.~(\ref{Gc}) in the form
\begin{widetext}
\begin{eqnarray}\label{GcK}
G^{c^2}_{d}(\nu;a_1,\ldots,a_d) &=&  \frac{
  \pi^{d/2}}{2^{\nu-d/2-1}\Gamma(\nu) \sqrt{a_1^2 \ldots a_d^2}}
\left[ 2^{\nu - d/2-1} \Gamma(\nu-d/2) c^{d-2\nu}  \right.  \nonumber
  \\ &+& \left.  \frac{2}{(2\pi)^{d/2-\nu}}  \sum_{i=1}^d
  \sum_{n_i=1}^{+\infty} \left(\frac{c a_i}{n_i}\right)^{d/2-\nu}
  K_{\nu-d/2}\left(\frac{2\pi  c n_i}{a_i}\right) \right.  \nonumber
  \\ &+& \left. \frac{2^d}{(2 \pi)^{d/2-\nu}} \sum_{n_1,\ldots ,
    n_d=1}^{\infty} \left(\frac{c}{\sqrt{ \frac{n_1^2}{a_1^2} + \ldots
      +\frac{n_d^2}{a_d^2}} }\right)^{d/2-\nu} K_{\nu-d/2}\left( 2 \pi
  c \sqrt{ \frac{n_1^2}{a_1^2} + \ldots +\frac{n_d^2}{a_d^2}} \right)
  \right],
\end{eqnarray} 
\end{widetext}
where $K_\alpha(x)$ is the modified Bessel function of the second
kind.

{}For definiteness, in this work we will focus in the case of
boundaries satisfying DBC.  With DBC, at the boundaries the fields
vanish (i.e., the fields should not ``leak" beyond the boundaries).
Hence, $\phi(x_i=0)=\phi(x_i=L)=\chi(x_i=0)=\chi(x_i=L) =0$, where
$x_i$ refers to those space directions suffering the compactification,
$x_i \in [0,L]$.  The discrete frequencies associated with the DBC are
then given by Eq.~(\ref{DBC}). 

Recalling that temperature is accounted for through a periodic
compactification (for bosons)  in Euclidean time and considering the
case of one compactified space dimension using DBC, with length $L$
(i.e., along this work we consider a slab geometry in the
three-dimensional space: $\mathbb{R}^2 \times [0,L]$ ), then, we have
that Eq.~(\ref{Gc}) changes to
\begin{equation}\label{G2}
G^{c^2}_{\rm DBC}(\nu;a_T,a_L)=\sum_{n \in \mathbb{Z}} \sum_{l_1 \in
  \mathbb{N}_{>0}}  \left(c^2+a_T^2 n^2 + a_L^2 l_1^2\right)^{-\nu},
\end{equation}
where $a_T = 2 \pi T$, $a_L = \pi/L$ and $\nu=\alpha - (D-2)/2$, with
$D=4-2\epsilon$.

Note that Eq.~(\ref{G2}) can also be written as
\begin{eqnarray}
G^{c^2}_{\rm DBC}(\nu;a_T,a_L)&=&-\frac{1}{2}\sum_{n \in \mathbb{Z}}
\left(c^2+a_T^2 n^2 \right)^{-\nu} \nonumber \\ &+& \frac{1}{2}
\sum_{n \in \mathbb{Z}} \sum_{l_1 \in \mathbb{Z}}  \left(c^2+a_T^2 n^2
+ a_L^2 l_1^2\right)^{-\nu}, \nonumber \\
\label{G22}
\end{eqnarray}
and the last term is then of the form of Eq.~(\ref{GcK}), while the
first term in  Eq.~(\ref{G22}), under analytic continuation with the
zeta-function method, is obtained by using the
result~\cite{Elizalde:1995hck}
\begin{widetext}
\begin{eqnarray}
\sum_{n \in \mathbb{Z}}   \left(c^2+a_T^2 n^2 \right)^{-\nu} &=&
\frac{\sqrt{\pi}}{a_T} \frac{\Gamma(\nu-1/2)}{\Gamma(nu)} c^{1-2\nu}
+ \frac{4 \pi^\nu}{\Gamma(\nu)} a_T^{-1/2-\nu} c^{1/2-\nu}
\sum_{n=1}^{\infty} n^{\nu-1/2}  K_{\nu-1/2} (2 \pi n c/a_T).
\label{aTterm}
\end{eqnarray}
Hence, in Eqs.~(\ref{MphiTL}) and (\ref{MchiTL}), the function
$I^{(1)}_2$ in the case of DBC is explicitly given by
\begin{eqnarray}
I^{(1),{\rm DBC}}_{2}(m_i,T,L) &=& -\frac{m_i^2}{16 \pi^2 \epsilon} +
\frac{m_i^2}{16 \pi^2} \left[ \ln \left(\frac{m_i^2}{\mu^2}
  \right)-1\right] +\frac{m_i}{8 \pi L}  + \frac{T}{4 \pi L} \ln
\left( 1- e^{ -m_i/T} \right) \nonumber \\ &+& \frac{m_i T}{2 \pi^2}
\sum_{n=1}^{+\infty} \frac{1}{n} K_1 \left(\frac{m_i}{T} n\right) +
\frac{m_i}{4 \pi^2 L}  \sum_{l_1=1}^{+\infty} \frac{1}{l_1} K_1
\left(2 m_i L l_1\right) \nonumber \\ &+& \frac{m_i}{\pi^2}
\sum_{n=1}^{+\infty}  \sum_{l_1=1}^{+\infty} \frac{1}{\sqrt{
    \frac{n^2}{T^2} + 4 l_1^2 L^2} } K_1\left( m_i \sqrt{
  \frac{n^2}{T^2} + 4l_1^2 L^2}  \right).
\label{I2RDBC}
\end{eqnarray}
\end{widetext}

The divergent term appearing in Eq.~(\ref{I2RDBC}) is the standard
divergence for the two-point Green function for a scalar field in the
bulk. Thus, as far as regularization and renormalization are
concerned, the (mass square) divergence can be removed by adding the
standard counterterm of mass in the vacuum. We note, however, that
more generally, when working with finite-size effects at the level of
the effective action, additional renormalization counterterms are
required as surface divergences also
appear~\cite{Fosco:1999rs,Caicedo:2002ft,Svaiter:2004ad}. In the
present work we will not have to deal with these more general
renormalization and, thus, we will only require the standard
renormalization counterterms in the bulk (see also
Ref.~\cite{Fucci:2017weg} for details on the renormalization under the
finite-size effects in general).  Thus, by adding to the  Lagrangian
density Eq.~(\ref{Lphich}) the appropriate mass counterterms of
renormalization, e.g., by redefining the masses of the fields such
that $m_\phi^2 \to m_\phi^2 + \delta m_\phi^2$ and $m_\chi^2 \to
m_\chi^2 + \delta m_\chi^2$, the mass counterterms are, respectively,
given by
\begin{eqnarray}
\delta m_\phi^2 = \frac{1}{32 \pi^2 \epsilon} \left( \lambda_\phi
\frac{N_\phi +2}{3} m_\phi^2 + \lambda N_\chi m_\chi^2 \right),
\label{mass-renorphi}
\end{eqnarray}
and
\begin{eqnarray}
\delta m_\chi^2 = \frac{1}{32 \pi^2 \epsilon} \left( \lambda_\chi
\frac{N_\chi +2}{3} m_\chi^2 + \lambda N_\phi m_\phi^2 \right).
\label{mass-renorchi}
\end{eqnarray}
The renormalized masses at one-loop order for the $\phi$ and $\chi$
fields then become
\begin{eqnarray}
M_\phi^2(T,L) &=& m_\phi^2 + \frac {\lambda_\phi}{2} \left ( \frac
{N_\phi+2}{3} \right ) I^{(1)}_{2,R}(m_\phi,T,L) \nonumber \\ &+&
\frac {\lambda}{2} N_\chi I^{(1)}_{2,R}(m_\chi,T,L),
\label{MphiTL}
\end{eqnarray}
and
\begin{eqnarray}
M_\chi^2(T,L) &=& m_\chi^2 + \frac {\lambda_\chi}{2} \left ( \frac
{N_\chi+2}{3} \right ) I^{(1)}_{2,R}(m_\chi,T,L) \nonumber \\ &+&\frac
{\lambda}{2} N_\phi I^{(1)}_{2,R}(m_\phi,T,L),
\label{MchiTL}
\end{eqnarray}
where $I^{(1)}_{2,R}(m_i,T,L)$ is the result obtained in
Eq.~(\ref{I2RDBC}) when the divergence is subtracted.

It is also illustrative to work in the approximation of small masses,
i.e., $m_i/T \ll 1$ and $m_i L \ll 1$. In this case, it is more
convenient to return to Eq.~(\ref{G2}).  Isolating the thermal zero
mode ($n=0$) from it, we have that
\begin{eqnarray}
\lefteqn{\sum_{n \in \mathbb{Z}} \sum_{l_1 \in  \mathbb{N}_{>0}}
  \left(m_i^2+a_T^2 n^2 + a_L^2 l_1^2\right)^{-\nu} } \nonumber \\ &&
=\sum_{l_1 \in  \mathbb{N}_{>0}} \left(m_i^2+ a_L^2
l_1^2\right)^{-\nu}  \nonumber \\ &&+ 2 \sum_{n \in \mathbb{N}_{>0}}
\sum_{l_1 \in  \mathbb{N}_{>0}}  \left(m_i^2+a_T^2 n^2 + a_L^2
l_1^2\right)^{-\nu}.
\label{G2smallm}
\end{eqnarray}
The first term in Eq.~(\ref{G2smallm}) can be written in a similar
form as in Eq.~(\ref{aTterm}), which, as also in according to
Ref.~\cite{Elizalde:1995hck}, can be expressed as
\begin{widetext}
\begin{eqnarray}
\sum_{l_1 \in  \mathbb{N}_{>0}} \left(m_i^2+ a_L^2 l_1^2\right)^{-\nu}
&=& -\frac{1}{2} (m_i^2)^{-\nu}  + \frac{\sqrt{\pi} \Gamma(\nu-1/2)}{2
  \Gamma(\nu)} a_L^{-2 \nu} \left(\frac{m_i^2}{a_L^2}\right)^{-\nu
  +1/2} \nonumber\\ &+& \frac{2 \pi^\nu}{\Gamma(\nu)}  a_L^{-2 \nu}
\left(\frac{m_i^2}{a_L^2}\right)^{-\nu/2 +1/4} \sum_{l_1=1}^{+\infty}
l_1^{\nu-1/2} K_{\nu-1/2} (2 \pi l_1 m_i/a_L).
\label{suml1}
\end{eqnarray}
\end{widetext}
The second term in Eq.~(\ref{G2smallm}) can now be written as
\begin{widetext}
\begin{eqnarray}
2 \sum_{n \in \mathbb{N}_{>0}} \sum_{l_1 \in  \mathbb{N}_{>0}}
\left(m_i^2+a_T^2 n^2 + a_L^2 l_1^2\right)^{-\nu} &=& 2 (2 \pi
T)^{-2\nu}  \sum_{n \in \mathbb{N}_{>0}} \sum_{l_1 \in
  \mathbb{N}_{>0}} \left( \frac{m_i^2}{4 \pi^2 T^2} + n^2 +
\frac{l_1^2}{4T^2 L^2} \right)^{-\nu}  \nonumber \\ &=& 2 (2 \pi
T)^{-2\nu}  \sum_{n \in \mathbb{N}_{>0}} \sum_{l_1 \in
  \mathbb{N}_{>0}} \left(  n^2 + \frac{l_1^2}{4T^2 L^2} \right)^{-\nu}
\nonumber \\ &-& 2 \nu \,(2 \pi T)^{-2\nu}   \frac{m_i^2}{4 \pi^2 T^2}
\sum_{n \in \mathbb{N}_{>0}} \sum_{l_1 \in  \mathbb{N}_{>0}} \left(
n^2 + \frac{l_1^2}{4T^2 L^2} \right)^{-1-\nu} + {\cal O}(m_i^4/T^4),
\label{moverTexpand}
\end{eqnarray}
\end{widetext}
where in the last part in Eq.~(\ref{moverTexpand}) we have expanded
for $m_i/T \ll 1$.

We can now use the  following result for the two-dimensional Epstein
zeta-function representation~\cite{Elizalde:1989mr}
\begin{widetext}
\begin{eqnarray}
\sum_{n_1,n_2=1}^{\infty}  \left(a_1 n_1^2 +a_2 n_2^2 \right)^{-s} &=&
-\frac{a_2^{-s}}{2} \zeta(2 s) + \frac{a_2^{-s}}{2} \left( \frac{\pi
  a_2}{a_1} \right)^{1/2} \frac{\Gamma(s-1/2)}{\Gamma(s)} \zeta(2 s-1)
\nonumber \\ &+& \frac{2 \pi^s}{\Gamma(s)} a_1^{-s/2-1/4}
a_2^{-s/2+1/4} \sum_{n_1,n_2=1}^{\infty}  n_1^{s-1/2} n_2^{-s+1/2}
K_{s-1/2} \left( 2 \pi \sqrt{\frac{a_2}{a_1}} n_1 n_2 \right).
\label{2dzeta}
\end{eqnarray}
\end{widetext}

In Eq.~(\ref{2dzeta}) we can choose $a_1$ and $a_2$ to be either $1/(2
T L)^2$ or $1$.  Obviously the two choices are completely equivalent,
however, in the cases where $m_i/T\ll 1$ and $T L > 1/2$, the choice
$a_1=1/(2 T L)^2$ and $a_2=1$ turns out to be the suitable one, as it
display better convergence properties in this case. Alternatively, we
could also make the expansion in terms of $m_i L \ll 1$, which then
the choice $a_1=1$ and $a_2=1/(2 T L)^2$ is the one that will exhibit
better convergence  properties when $T L < 1/2$. Given this, in
practice, in all our numerical results,  we work with an interpolated
version of these expressions, favoring one case or the other depending
on the value of $T L$. This strategy is similar to the one used in
Ref.~\cite{Mogliacci:2018oea}.

Using the above expressions, we then obtain the small masses
expansions for Eq.~(\ref{I2RDBC}) and, when $|m_i|/T \ll 1$ and $T L
\geq 1/2$ and keeping terms up to quadratic order in the masses, it is
explicitly given by
\begin{eqnarray}
I^{(1),{\rm DBC}}_{2}(m_i,T,L) &\simeq&  -\frac{m_i^2}{16 \pi^2
  \epsilon}  \nonumber \\ &-& \frac{T}{2 \pi L} \ln \left[ \eta(2 i T
  L) \right]- \frac{m_i T}{4 \pi}  \nonumber \\ &+& \frac{T}{4 \pi L}
\left[ \ln\left( \frac{m_i}{T} \right) -\ln\left(1-e^{-2 m_i L}\right)
  \right] \nonumber \\ &+& \frac{m_i^2}{8 \pi^2} \left\{
\ln\left(\frac{4 \pi T}{\mu} \right) - \gamma_E  +  \frac{\pi T L}{3}
\right.  \nonumber \\ &+& \left. \frac{\pi}{12 T L} + 2 \ln \left[
  \eta(2 i T L) \right]  \right\},
\label{I2RDBCsmallm1}
\end{eqnarray}
while for $|m_i| L \ll 1$ and  $T L < 1/2$, we have that
\begin{eqnarray}
I^{(1),{\rm DBC}}_{2}(m_i,T,L) &\simeq&  -\frac{m_i^2}{16 \pi^2
  \epsilon}  \nonumber \\
&-& \frac{T}{2 \pi L} \ln \left( \eta[i/(2 T
  L)] \right) - \frac{m_i T}{4 \pi}  \nonumber \\ &+& \frac{T}{4 \pi
  L} \left[ \ln\left( 2 m_i L \right) -\ln\left(1-e^{-2 m_i L}\right)
  \right] \nonumber \\
&+& \frac{m_i^2}{8 \pi^2} \left\{
\ln\left(\frac{2 \pi}{L \mu} \right) - \gamma_E  +  \frac{\pi T L}{3}
\right.  \nonumber \\ &+& \left. \frac{\pi}{12 T L} + 2 \ln \left(
\eta[i/(2 T L)] \right)  \right\},
\label{I2RDBCsmallm2}
\end{eqnarray}
where in the above expressions $\eta(x)$ is the Dedekind eta function,
defined as~\cite{apostol}
\begin{equation}
\eta(x) = e^{i \pi x/12} \prod_{n=1}^\infty \left( 1- e^{2 \pi i n x}
\right).
\end{equation}

As a cross-check of Eq.~(\ref{I2RDBCsmallm1}), let us consider the bulk
limit $L\to \infty$ of it.  Using the identities,
\begin{eqnarray}
\lim_{x\to \infty} \frac{\ln[\eta(2 i x)]}{2 x} = -\frac{\pi}{12}, 
\label{ide1}
\end{eqnarray}
and
\begin{eqnarray}
\lim_{x\to \infty}\left\{ \frac{\pi x}{6} + \ln[\eta(2 i x)] \right\}
= 0,
\label{ide2}
\end{eqnarray}
then Eq.~(\ref{I2RDBCsmallm1}) leads to the high-temperature
approximation, $m_i/T \ll 1$,
\begin{eqnarray}
I^{(1),{\rm DBC}}_{2}(m_i,T,L)\Bigr|_{L\to \infty} &=&
-\frac{m_i^2}{16 \pi^2 \epsilon} + \frac{T^2}{12}  - \frac{m_i T}{4
  \pi} \nonumber \\ &+& \frac{m_i^2}{8 \pi^2} \left[ \ln\left(\frac{4
    \pi T}{\mu} \right) - \gamma_E \right]  \nonumber \\ &+&{\cal
  O}(m_i^4/T^4),
\label{highTI1}
\end{eqnarray}
which is the correct expression for the thermal integral $I^{(1)}$ in
the high-temperature approximation~\cite{Laine:2016hma}.

It is also useful to work the limiting case where Eq.~(\ref{I2RDBCsmallm2}) applies.
Considering now the different dimensional parameters satisfy $m \ll T \ll 1/L$ and applying
again the identities  (\ref{ide1}) and (\ref{ide2}) now to Eq.~(\ref{I2RDBCsmallm2}), we obtain
\begin{eqnarray}
I^{(1),{\rm DBC}}_{2}(m_i,T,L)\Bigr|_{m \ll T \ll 1/L} &=&
-\frac{m_i^2}{16 \pi^2 \epsilon} + \frac{1}{48 L^2} \nonumber \\ &+& \frac{m_i^2}{8 \pi^2} \left[ \ln\left(\frac{2
    \pi}{L \mu} \right) - \gamma_E \right]  \nonumber \\ &+&{\cal
  O}(m_i^2L^{-2}/T^2).
\label{lowLI1}
\end{eqnarray}

\begin{center}
\begin{figure}[htb]
\subfigure[]{\includegraphics[width=7cm]{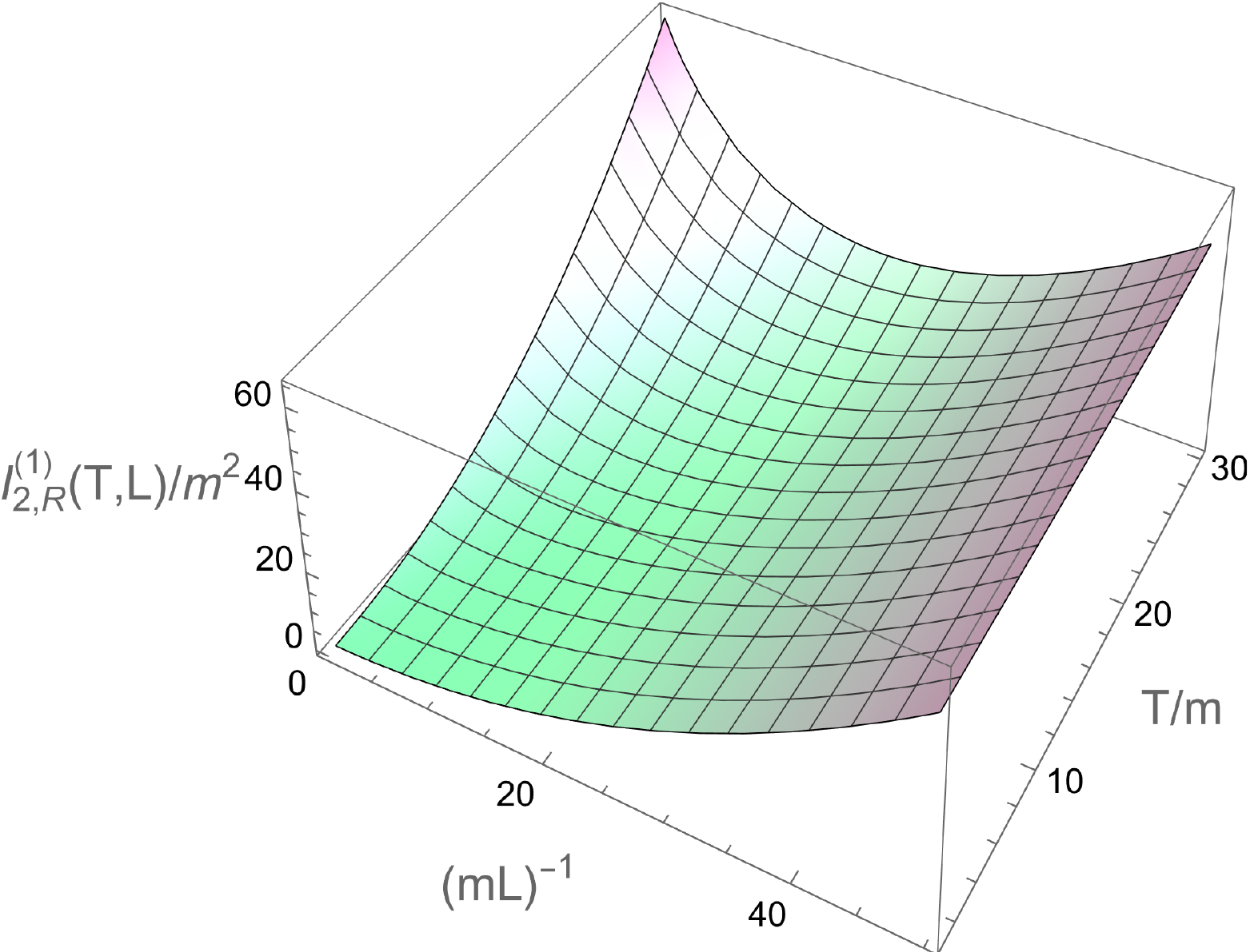}}
\subfigure[]{\includegraphics[width=7cm]{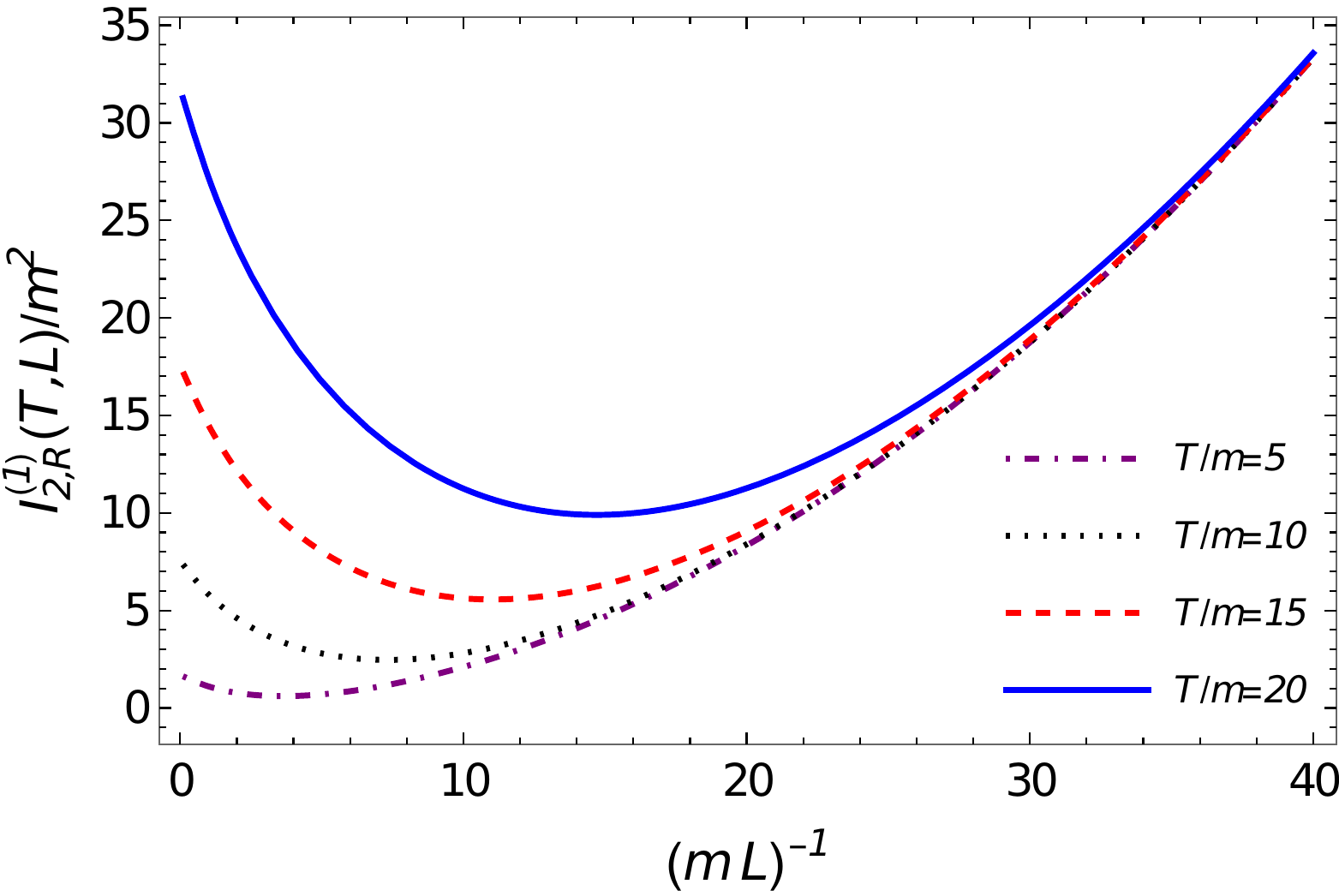}}
\caption{Panel (a): The variation of the renormalized two-point function with $T$ and $L$.
Panel (b), the same but for fixed values of the temperature. We have assumed $\mu=m$ for illustration. }
\label{fig1}
\end{figure}
\end{center}

The limiting case shown in Eq.~(\ref{highTI1}), by dropping the divergent and mass dependent terms, 
leads automatically to the perturbative quadratic mass as given by Eq.~(\ref{MiT}). On the other hand,
for  $m \ll T \ll 1/L$, we get similarly that
\begin{equation}
M_{i}^2(L) \simeq m^2_i + \frac{1}{48 L^2}  \left(\frac{N_{i}+2}{6} \,
\lambda_{i} +  \frac{N_{j}}{2} \lambda \right), \;\;{i,j=\phi,\chi}.
\label{MiL}
\end{equation}
Hence, the roles played by $T$ and $L$ get reversed. Therefore,
we expect that for sufficiently small values for $L$, an initially broken
symmetry will get restored  whenever
$(N_{i}+2)\lambda_{i} +  3N_{j} \lambda >0$. Otherwise, if $(N_{i}+2)\lambda_{i} +  3N_{j} \lambda <0$,
but still satisfying the boundness, Eq.~(\ref{boundness}), when $m_i^2 >0$,
ISB will emerge, while for  $m_i^2 <0$ we can have SNR.
More interestingly, since by keeping $L$ fixed and for large $T$ we have that
$I^{(1),{\rm DBC}}_{2,{\rm R}} \sim T^2/12$, while keeping $T$ fixed and small $L$
we have instead $I^{(1),{\rm DBC}}_{2,{\rm renor}} \sim 1/(48 L^2)$, it is expected
that the thermal two-point function will necessarily have a minimum for some ranges
of $T$ and $L$ values. This is in fact confirmed by a plot of the renormalized function
$I^{(1),{\rm DBC}}_{2}(m_i,T,L)$ in {}Fig.~\ref{fig1}.

As an immediate consequence of the behavior of $I^{(1),{\rm DBC}}_{2,{\rm R}}$ as a function of $T$ and $L$
will be the possibility of having 
behavior analogous to reentrant phase transitions as we will demonstrate in Sec.~\ref{results}.

\section{Renormalized masses and couplings and resummation procedure}
\label{section4}

Let us discuss in this section both the dependence of the renormalized masses and couplings with the
scale and also the resummation procedure we have adopted to analyze the phase transition patterns 
at both finite temperature and size.

\subsection{Dependence of the masses and couplings with the renormalization scale}
\label{scales}

Before presenting our results, let us discuss the dependence of the masses and coupling
in our model with the renormalization scale $\mu$.

As already mentioned in the previous section, it is enough for us here to work the renormalization
of the system in the bulk. Working at the one-loop level, the renormalized masses and coupling constants
can then be readily computed (see, for instance Ref.~\cite{Kleinert:2001hn})
and they are given by
\begin{eqnarray}
m_\phi^2(\mu) &=& \bar{m}_\phi^2 + \frac {\bar{\lambda}_\phi}{2} \left( \frac
      {N_\phi+2}{3} \right) \frac{\bar{m}_\phi^2}{16 \pi^2} 
\left[ -1 + \ln\left(\frac{\bar{m}_\phi^2}{\mu^2}\right) \right]
      \nonumber \\ &+&\frac {\bar{\lambda}}{2} N_\chi
     \frac{\bar{m}_\chi^2}{16 \pi^2} 
\left[ -1 + \ln\left(\frac{\bar{m}_\chi^2}{\mu^2}\right) \right],
\label{mphimu}
\\ 
m_\chi^2(\mu) &=& \bar{m}_\chi^2 + \frac {\bar{\lambda}_\chi}{2} \left( \frac
      {N_\chi+2}{3} \right) \frac{\bar{m}_\chi^2}{16 \pi^2} 
\left[ -1 + \ln\left(\frac{\bar{m}_\chi^2}{\mu^2}\right) \right]
      \nonumber \\ &+&\frac {\bar{\lambda}}{2} N_\phi
     \frac{\bar{m}_\phi^2}{16 \pi^2} 
\left[ -1 + \ln\left(\frac{\bar{m}_\phi^2}{\mu^2}\right) \right],
\label{mchimu}
\\
\lambda_{\phi}(\mu) &=& \bar{\lambda}_\phi + \frac{(N_\phi + 8)\bar{\lambda}_\phi^2}{6(4\pi)^2}  
\ln\left(\frac{\bar{m}_\phi^2}{\mu^2}\right) 
\nonumber\\
&+& \frac{3N_\chi \bar{\lambda}^2}{2} \ln\left(\frac{\bar{m}_\chi^2}{\mu^2}\right),
\label{lphimu}
\\
\lambda_{\chi}(\mu) &=& \bar{\lambda}_\chi + \frac{(N_\chi + 8)\bar{\lambda}_\chi^2}{6(4\pi)^2}  
\ln\left(\frac{\bar{m}_\chi^2}{\mu^2}\right) 
\nonumber\\
&+&  \frac{3N_\phi \bar{\lambda}^2}{2} \ln\left(\frac{\bar{m}_\phi^2}{\mu^2}\right),
\label{lchimu}
\\
\lambda(\mu) &=& \bar{\lambda} + 2 \bar{\lambda}^2  \frac{1}{\bar{m}_\phi^2-\bar{m}_\chi^2}\left\{
 \frac{\bar{m}_\phi^2}{16 \pi^2}\left[ -1 + \ln\left(\frac{\bar{m}_\phi^2}{\mu^2}\right) \right]
\right.
\nonumber \\
&-& \left. \frac{\bar{m}_\chi^2}{16 \pi^2} 
\left[ -1 + \ln\left(\frac{\bar{m}_\chi^2}{\mu^2}\right) \right]
\right\}\nonumber\\
&+& \frac{(N_\phi + 2)\bar{\lambda} \bar{\lambda}_\phi}{6(4\pi)^2}  
\ln\left(\frac{\bar{m}_\phi^2}{\mu^2}\right) 
\nonumber \\
&+&  \frac{(N_\chi + 2)\bar{\lambda} \bar{\lambda}_\chi}{6(4\pi)^2}  
\ln\left(\frac{\bar{m}_\chi^2}{\mu^2}\right),
\label{lmu}
\end{eqnarray}
with the above equations depending on the renormalization scale $\mu$
through the $\overline{\rm MS}$ parameters of the renormalized
Lagrangian.  
Note that in the above equations we are expressing all bare quantities (which does not depend
on the scale $\mu$), and that appears on the right-hand side of the equations, as a barred quantity. 
The renormalized parameters, on the other hand, are all dependent on the scale $\mu$.

{}Finally, note that the above equations apply to the symmetry restored case where $\bar{m}_i^2 >0$.
In the symmetry broken case, the procedure is well-known (see, for instance 
Refs.~\cite{Nishijima:1979yx,Taylor:1983xu,Brown:1988yj} for details of the corresponding renormalization
group (RG) procedure):
the fields are shifted around the true mininum, leading to bare and renormalized masses, when in the
vacuum, replaced by $\bar{m}_i^2 \to 2 |\bar{m}_i^2|$ and $m_i^2 \to 2 |m_i^2|$, respectively.

The renormalization scale dependence on both masses and couplings are
given in terms of their  respective RG
expressions, given in terms of the RG functions $\beta$ and $\gamma_m$
given, respectively, by~\cite{Kleinert:2001hn}
\begin{eqnarray}
\beta_i = \mu \frac{\partial \lambda_i}{\partial \mu},
\label{RGbeta}
\end{eqnarray}\
and
\begin{eqnarray}
\gamma_{m_i} = \mu \frac{\partial \ln m_i}{\partial \mu},
\label{RGgammam}
\end{eqnarray}
which, for our respective masses and couplings and at one-loop
($\hbar$) order, give
\begin{eqnarray}
\gamma_{m_\phi} = \frac{1}{2 m_\phi^2(\mu) (4 \pi)^2 } &&\left[
  \lambda_\phi(\mu) \frac{(N_\phi+2)}{3} m_\phi^2(\mu) \right.
  \nonumber \\ &&+\left. \lambda(\mu) N_\chi m_\chi^2(\mu)
  \frac{}{}\right],
\label{gammamphi}
\end{eqnarray}
\begin{eqnarray}
\gamma_{m_\chi} = \frac{1}{2 m_\chi^2(\mu) (4 \pi)^2 } &&\left[
  \lambda_\chi(\mu) \frac{(N_\chi+2)}{3} m_\chi^2(\mu) \right.
  \nonumber \\ &&+ \left. \lambda(\mu) N_\phi m_\phi^2(\mu) \right],
\label{gammamchi}
\end{eqnarray}
\begin{eqnarray}
&&\beta_{\lambda_\phi}=\frac{(N_\phi+8)}{3}
  \frac{\lambda_\phi^2(\mu)}{(4\pi)^2} + 3N_\chi
  \frac{\lambda^2(\mu)}{(4\pi)^2},
\label{betaphi}
\end{eqnarray}
\begin{eqnarray}
&&\beta_{\lambda_\chi}=\frac{(N_\chi+8)}{3}
\frac{\lambda_\chi^2(\mu)}{(4\pi)^2} + 3N_\phi
\frac{\lambda^2(\mu)}{(4\pi)^2},
\label{betachi}
\end{eqnarray}
\begin{eqnarray}
\beta_{\lambda}&=& 4 \frac{\lambda^2(\mu)}{(4\pi)^2}+
\frac{(N_\phi+2)}{3} \frac{\lambda(\mu)\lambda_\phi(\mu)}{(4\pi)^2}
\nonumber \\ &+&\frac{(N_\chi+2)}{3}
\frac{\lambda(\mu)\lambda_\chi(\mu)}{(4\pi)^2}.
\label{beta}
\end{eqnarray}

Equations~(\ref{gammamphi})-(\ref{beta}) form a coupled
set of flow equations determining how the masses and couplings change
under different renormalization scales. In particular, we can also see
from, e.g., Eqs.~(\ref{betaphi}), (\ref{betachi}) and (\ref{beta}),
that their solution is also equivalent to solving the coupled set of
linear equations,
\begin{eqnarray}
\lambda_{\phi}(\mu) &=& \lambda_{\phi}(\mu_0) + \frac {1}{(4 \pi)^2}
\ln \left ( \frac {\mu}{\mu_0} \right ) \nonumber \\ &\times& \left[
  \frac {(N_{\phi}+8)}{3} \lambda_{\phi}(\mu_0) \lambda_\phi(\mu) +
  3N_{\chi} \lambda(\mu_0) \lambda(\mu) \right ],  \nonumber \\
\label{setphi}
\\ \lambda_{\chi}(\mu) &=& \lambda_{\chi}(\mu_0) + \frac {1}{(4
  \pi)^2} \ln \left ( \frac {\mu}{\mu_0} \right )
\nonumber
\\ &\times& \left[  \frac {(N_{\chi}+8)}{3} \lambda_{\chi}(\mu_0)
  \lambda_\chi(\mu) +  3N_{\phi} \lambda(\mu_0) \lambda(\mu) \right ],
\nonumber \\
\label{setchi} 
\\ \lambda(\mu) &=& \lambda(\mu_0)  + \frac {\lambda(\mu_0)}{(4\pi)^2}
\ln \left ( \frac {\mu}{\mu_0} \right )  \nonumber \\ &\times& \left [
  \frac {1}{2}  \left ( \frac {N_{\phi}+2}{3} \right )
  \lambda_{\phi}(\mu) + \frac {1}{2} \left ( \frac {N_{\chi}+2}{3}
  \right )  \lambda_{\chi}(\mu) \right ] \nonumber \\ &+& \frac
    {\lambda(\mu)}{(4\pi)^2}  \ln \left ( \frac {\mu}{\mu_0} \right )
    \nonumber \\
    &\times& \left [  \frac {1}{2}  \left ( \frac
      {N_{\phi}+2}{3} \right )  \lambda_{\phi}(\mu_0) + \frac {1}{2}
      \left ( \frac {N_{\chi}+2}{3} \right )  \lambda_{\chi}(\mu_0)
      \right ]
    \nonumber \\ & +& \frac {1}{4 \pi^2} \ln \left ( \frac
              {\mu}{\mu_0} \right ) \lambda(\mu_0) \lambda(\mu),
\label{setlambda}
\end{eqnarray}
which also make apparent how the renormalized couplings are related
through a change of the scale from, e.g., $\mu_0$ to $\mu$.  The
results obtained from the flow equations given above give the standard
way of nonperturbatively resumming through the RG equations the
leading order corrections (in this case the leading log-dependent
corrections) to the coupling constants.  These equations also show
that the couplings evolve with the scale in a logarithmic
way. Suitable choices of the scale can minimize these logarithmic
contributions. It is common in the literature in general that at high
temperatures to take the scale proportional to the temperature, $\mu
\sim T$. In particular, a suitable choice of scale has been shown to
be given by~\cite{Drummond:1997cw} $\mu = 2 \pi T$.  

The typical strategy is followed: for example, in the hard-thermal loop analysis
of the thermodynamics (see, for example, Ref.~\cite{Andersen:2004fp}),
we can express all physical quantities in terms of the renormalized, scale-dependent, 
parameters. Working similarly here, this means, for instance,
the set of renormalized parameters are given at the reference scale
$\mu_0$. By a change of the scale (e.g., with the temperature, or, in our present problem
can also be with $L$), the new set of renormalized parameters related to the
original ones fixed at the reference scale $\mu_0$ are then obtained by solving
the coupled flow equations,  Eqs.~(\ref{gammamphi})-(\ref{beta}), for the new value $\mu$.
The bare parameters are then finally determined by inverting the set of
equations (\ref{mphimu})-(\ref{lmu}), getting
$\bar{m}_\phi(m_\phi,m_\chi,\lambda_\phi,\lambda_\chi,\lambda)$, etc.

At finite sizes,
we also see from the equations derived earlier,
Eqs.~(\ref{I2RDBCsmallm1}) and (\ref{I2RDBCsmallm2}), that those
equations suggest that a more suitable scale might be $\mu \sim 1/L$
when $T L < 1/2$. In practice, in all of our numerical analysis, we
will adopt for the reference scale $\mu_0=2 \pi T$ or $\mu_0 = 2 \pi/L$, depending whether the
temperature or the (inverse of the) length size dominates the loop
integrals. As typically adopted in the literature to check the variation of the results
with the scale $\mu$, we will then vary $\mu$ in a range $\alpha = \mu/\mu_0= [1/2, 2]$.
Either way, the logarithmic dependence on the scale will
imply that all of our results will be weakly dependent on the
particular choice of $\mu$, provided that the couplings, $T$ and $1/L$ are not too large. 
In the parameters we will be working with our examples in Sec.~\ref{results}, this will
always be the case. 

\subsection{The bubble (ring) resummed gap equations for the masses and field expectation values}
\label{secring}

We want to investigate the full phase structure for the multiple field
system. However, it is well known that PT breaks down
at high temperatures and, in particular close to critical
points~\cite{Bellac:2011kqa}.  Here, we go beyond the perturbation
theory by using the bubble (or ring) dressed method for finite-temperature 
scalar fields~\cite{Espinosa:1992gq}. In the bubble
resummation procedure, the one-loop terms with the temperature and
finite-size effects are self-consistently resummed by using the gap
equations for the masses, which is similar as previously considered in
ISB/SNR earlier
studies~\cite{Roos:1995vm,Pinto:2005ey,Pinto:2006cb,Phat:2007zz}. In
this case, this is equivalent of solving the coupled system of gap
equations for the renormalized masses\footnote{Note that these
  equations can also be seen as the extension to multiple fields of
  the foam diagram resummation defined, e.g., in
  Ref.~\cite{Drummond:1997cw}.},
\begin{eqnarray}
M_\phi^2 &=& \bar{m}_\phi^2 + \frac {\bar{\lambda}_\phi}{2} \left ( \frac
      {N_\phi+2}{3} \right ) I^{(1)}_{2,R}(M_\phi,T,L)
      \nonumber \\ &+&\frac {\bar{\lambda}}{2} N_\chi
      I^{(1)}_{2,R}(M_\chi,T,L),
\label{MphiTLgap}
\\ M_\chi^2 &=& \bar{m}_\chi^2 + \frac {\bar{\lambda}_\chi}{2} \left (
\frac {N_\chi+2}{3} \right )  I^{(1)}_{2,R}(M_\chi,T,L)
\nonumber \\ &+& \frac {\bar{\lambda}}{2} N_\phi
I^{(1)}_{2,R}(M_\phi,T,L),
\label{MchiTLgap}
\end{eqnarray}
where the renormalized one-loop function  $I^{(1)}_{2,R}$ is given by
Eq.~(\ref{I2RDBC}) with the divergence subtracted.

The fields are shifted around their expectation values, $\phi \to \bar
\phi + \phi'$ and $\chi \to \bar \chi + \chi'$, where $\bar \phi
\equiv \langle \phi \rangle$ and $\bar \chi \equiv \langle \chi
\rangle$, while $\langle \phi' \rangle=0$ and $ \langle \chi'
\rangle=0$.  The expectation values for the fields  are then defined
through their respective coupled tadpole equations, given by,
respectively, in terms of the one-particle irreducible (1PI) one-point
functions $\Gamma^{(1)}_\phi\equiv 0$ and $\Gamma^{(1)}_\chi\equiv
0$. At the one-loop level and resummed bubble approximation,  we have
that
\begin{eqnarray}
\Gamma^{(1)}_\phi &=& - \bar{m}_\phi^2 \bar\phi - \frac{\bar{\lambda}_\phi}{6}
\bar\phi^3 - \frac{\bar{\lambda}}{2}\bar\phi \bar \chi^2 \nonumber \\ &-&
\frac{(N_\phi+2)}{6} \bar{\lambda}_\phi \bar\phi
I^{(1)}_{2,R}(\tilde{\Omega}_\phi,T,L) \nonumber \\ &-&\frac
{N_\chi}{2} \bar{\lambda} \bar\phi I^{(1)}_{2,R}(\tilde{\Omega}_\chi,T,L)
\equiv 0,
\label{Gammaphi}
\\ \Gamma^{(1)}_\chi &=& - \bar{m}_\chi^2 \bar\chi - \frac{\bar{\lambda}_\chi}{6}
\bar\chi^3 - \frac{\bar{\lambda}}{2} \bar \phi^2 \bar\chi\nonumber \\ &-&
\frac{(N_\chi+2)}{6} \bar{\lambda}_\chi \bar\chi
I^{(1)}_{2,R}(\tilde{\Omega}_\chi,T,L) \nonumber \\ &-&\frac
{N_\phi}{2} \bar{\lambda} \bar\chi
I^{(1)}_{2,R}(\tilde{\Omega}_\phi,T,L)\equiv 0,
\label{Gammachi}
\end{eqnarray}
where $\tilde{\Omega}_\phi$ and $\tilde{\Omega}_\chi$ are defined,
respectively, by
\begin{eqnarray}
\tilde{\Omega}_\phi^2 &=& M_\phi^2 + \frac{\bar{\lambda}_\phi}{2}
\bar\phi^2 + \frac{\bar{\lambda}}{2} \bar\chi^2, \\ \tilde{\Omega}_\chi^2
&=& M_\chi^2 + \frac{\bar{\lambda}_\chi}{2} \bar\chi^2 +
\frac{\bar{\lambda}}{2} \bar\phi^2.
\end{eqnarray}

We now have the complete set of necessary equations to analyze the
phase structure of the coupled two-scalar field system.

\section{Results}
\label{results}

Let us now consider the combined effects of temperature and finite
size in the phase transition patterns in the two-scalar field
model. But before studying the  coupled two-field system, it is
instructive to first analyze the case of only one field, e.g., $\phi$,
which can be easily obtained by setting the  intercoupling $\lambda$
to zero in the equations defined in the previous section.

\begin{center}
\begin{figure*}[htb]
\subfigure[]{\includegraphics[width=7cm]{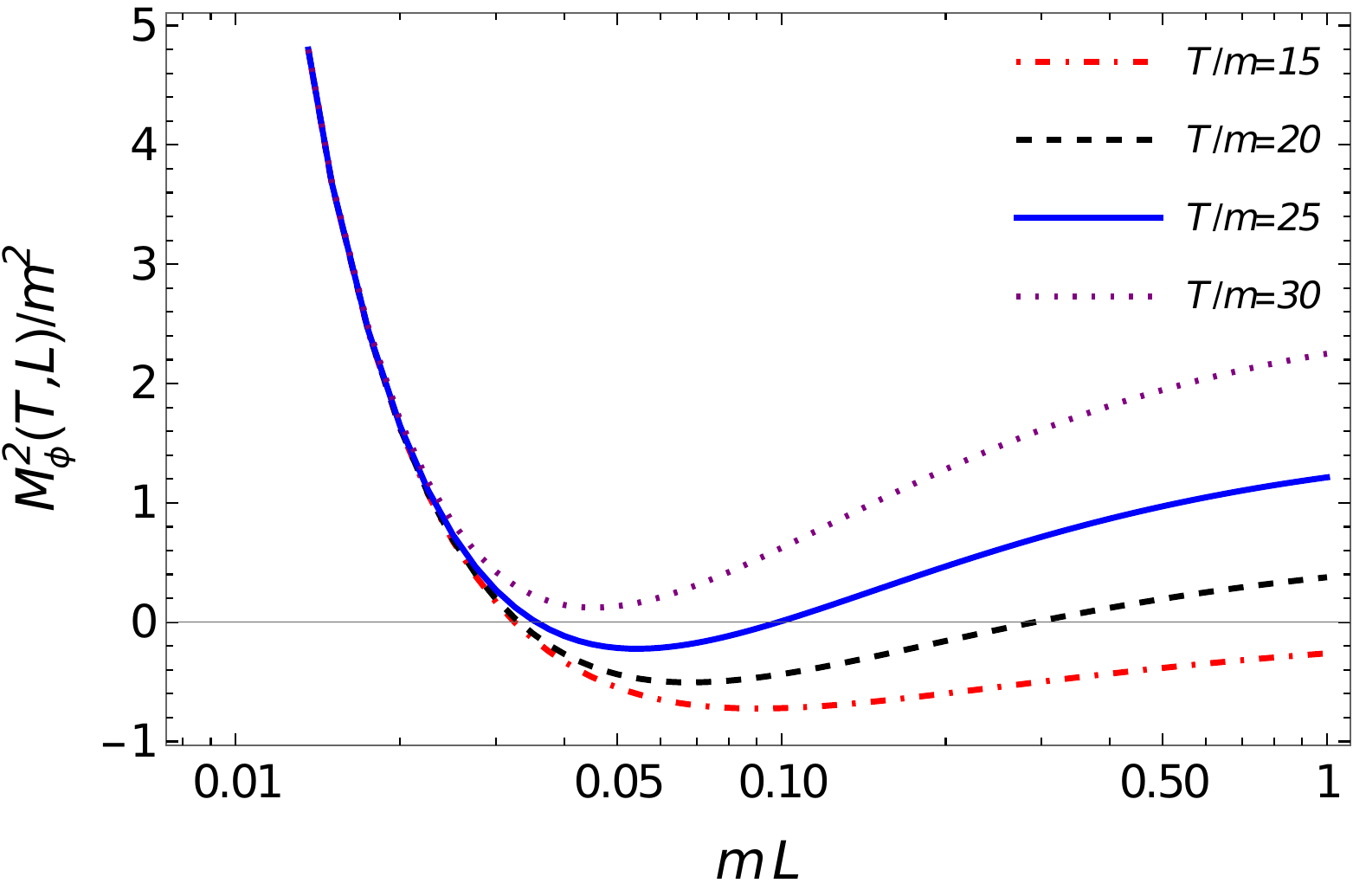}}
\subfigure[]{\includegraphics[width=7cm]{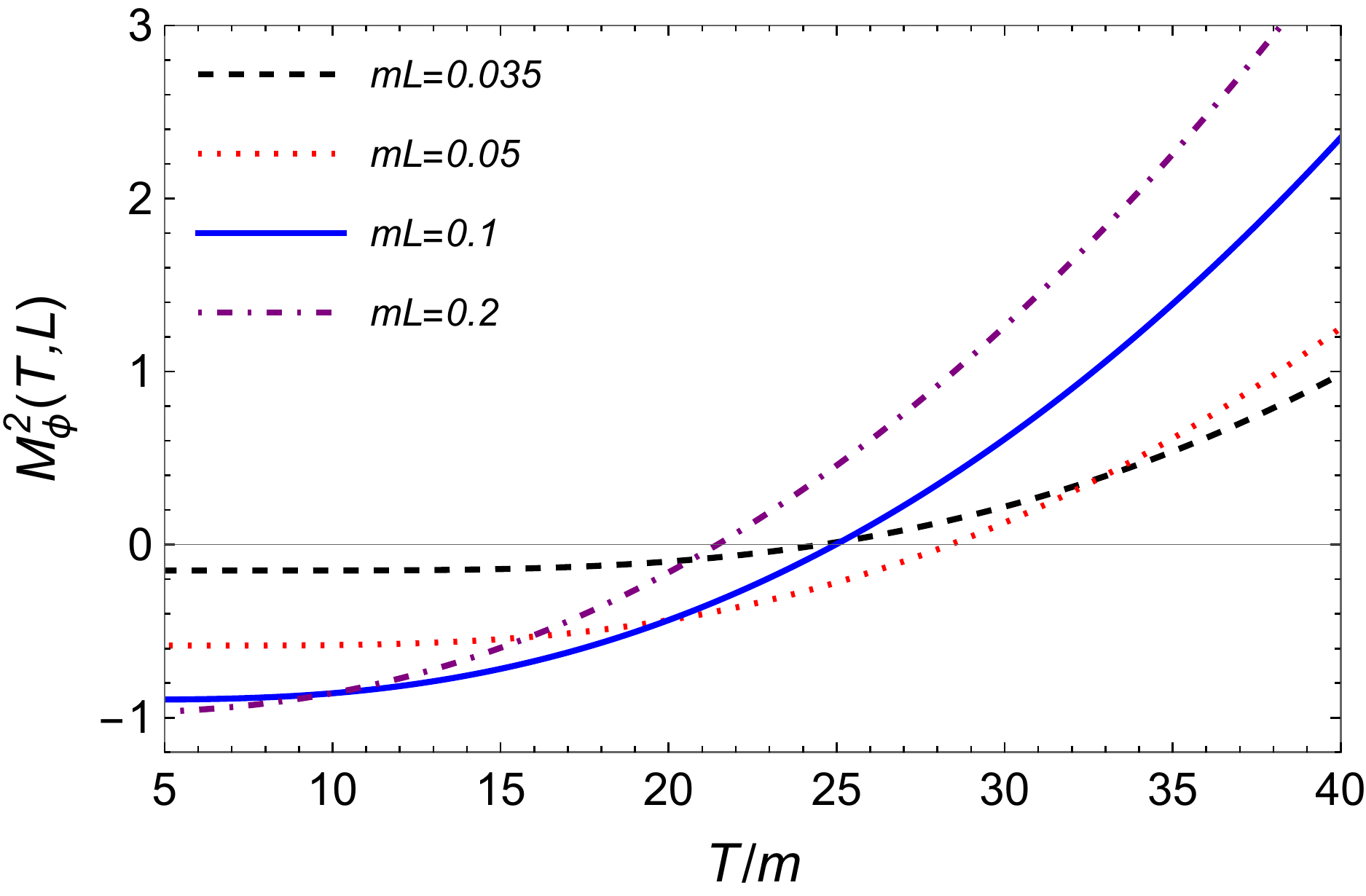}}
\subfigure[]{\includegraphics[width=7cm]{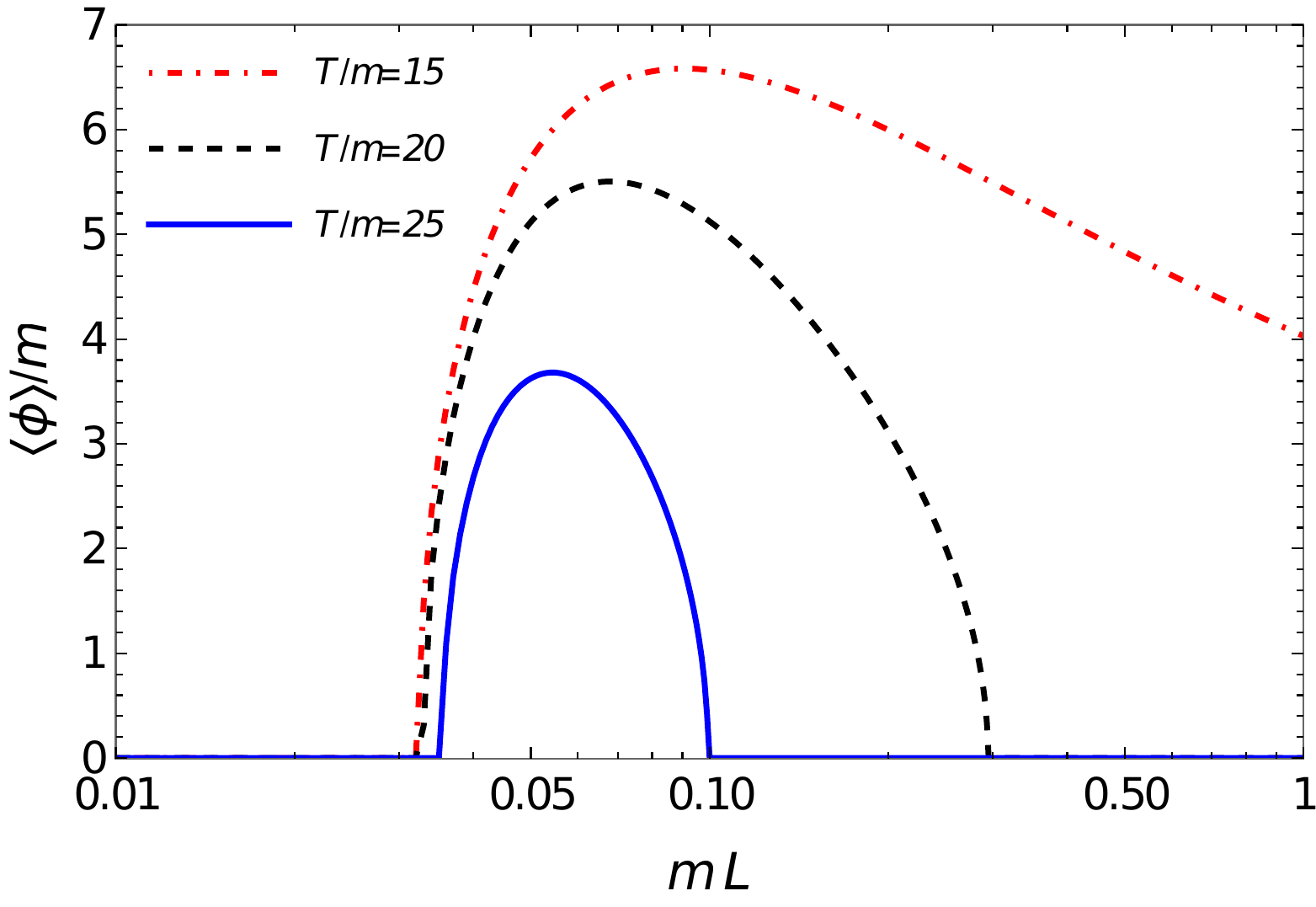}}
\subfigure[]{\includegraphics[width=7cm]{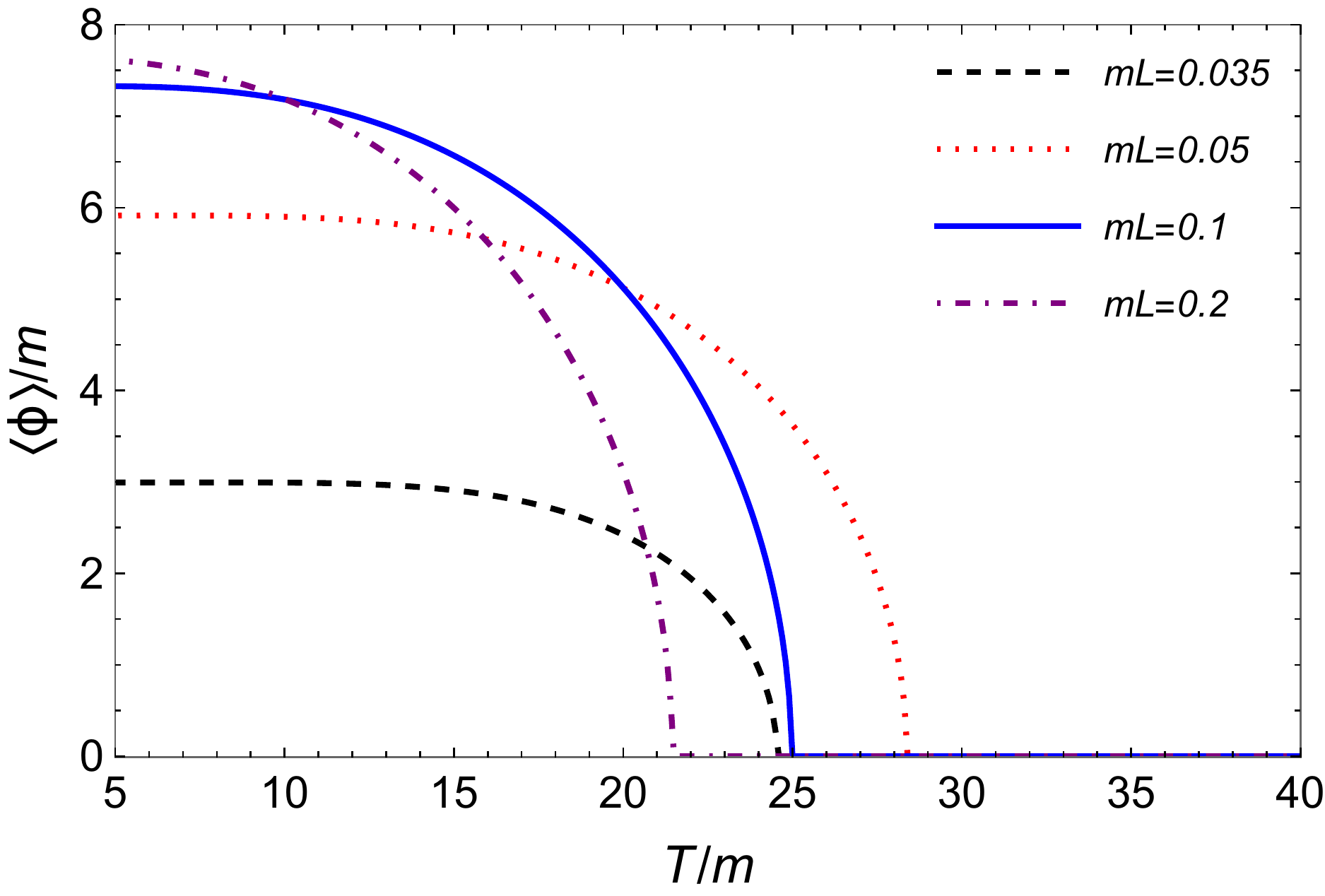}}
\caption{The effective gap mass squared for $\phi$ and the expectation
  value $\langle \phi \rangle$ in the one-field case ($\lambda=0$) 
in the case of a symmetry breaking tree-level potential as
  a function of the length $L$ for fixed values of temperature (panels
  a and c) and as a function of the temperature $T$ for fixed values
  of length $L$ (panels b and d). The renormalization scale is here fixed at the 
reference scale, $\mu =\mu_0$ (see Sec.~\ref{scales}), the renormalized 
coupling constant is $\lambda_\phi(\mu_0)=0.1$, we have also considered  $N_\phi=1$
and all dimensional quantities are in units of the renormalized mass $m\equiv m(\mu_0)$. 
}
\label{fig2}
\end{figure*}
\end{center}

\subsection{The one-scalar field case}

Setting the tree-level mass square term in the potential as negative,
i.e., $\bar{m}_\phi^2 = -|\bar{m}_\phi|^2\equiv - m^2$, in {}Fig.~\ref{fig2} we
show the effective mass $M_\phi(T,L)$ and the field expectation value
$\langle \phi \rangle$ in the DBC as a function of $L$ for fixed
temperature values [panels (a) and (c)] and as a function of the
temperature for fixed values of $L$ [panels (b) and (d)].

{}From {}Figs.~\ref{fig2}(a) and \ref{fig2}(c) we can see that the
effect of the compact dimension allows the system to have a double
critical point, where the symmetry can get broken in between two
values of critical length $L_c$. This is a reentrant phase transition,
which might be of interest in condensed matter
systems~\cite{Pinto:2005ey,Ramos:2006er,Pinto:2006cb,Linhares:2012vr}. These
type of transitions are also of interest for understanding, for
example, phase transitions in superconducting films, as studied
previously in Ref.~\cite{Linhares:2006my}, which considered finite-size 
effects in a  Ginzburg–Landau model with periodic boundary
conditions. The same trend of transitions also is seen to happen here
in the context of DBC. In {}Figs.~\ref{fig2}(b) and \ref{fig2}(d) the
SR behavior as a function of temperature is shown
for some fixed values of length $L$. 
Here we note that although there
seems to be no multiple critical points in the temperature (e.g., the
appearance of two critical values of temperature), it shows instead
that for small values of $L$ the critical temperature tends to grow as
$L$ increases, but then $T_c$ starts to decrease beyond some value of
$L$. Note that the phase transition points agree between the results
shown for either the effective gap mass or the field expectation
value, as they should. We can also notice from the results displayed
in the panels of {}Fig.~\ref{fig2} that the transition is always
continuous, which characterize a second-order phase transitions for
all the cases displayed in {}Fig.~\ref{fig2}. 

\begin{center}
\begin{figure*}[htb]
\subfigure[]{\includegraphics[width=5.7cm]{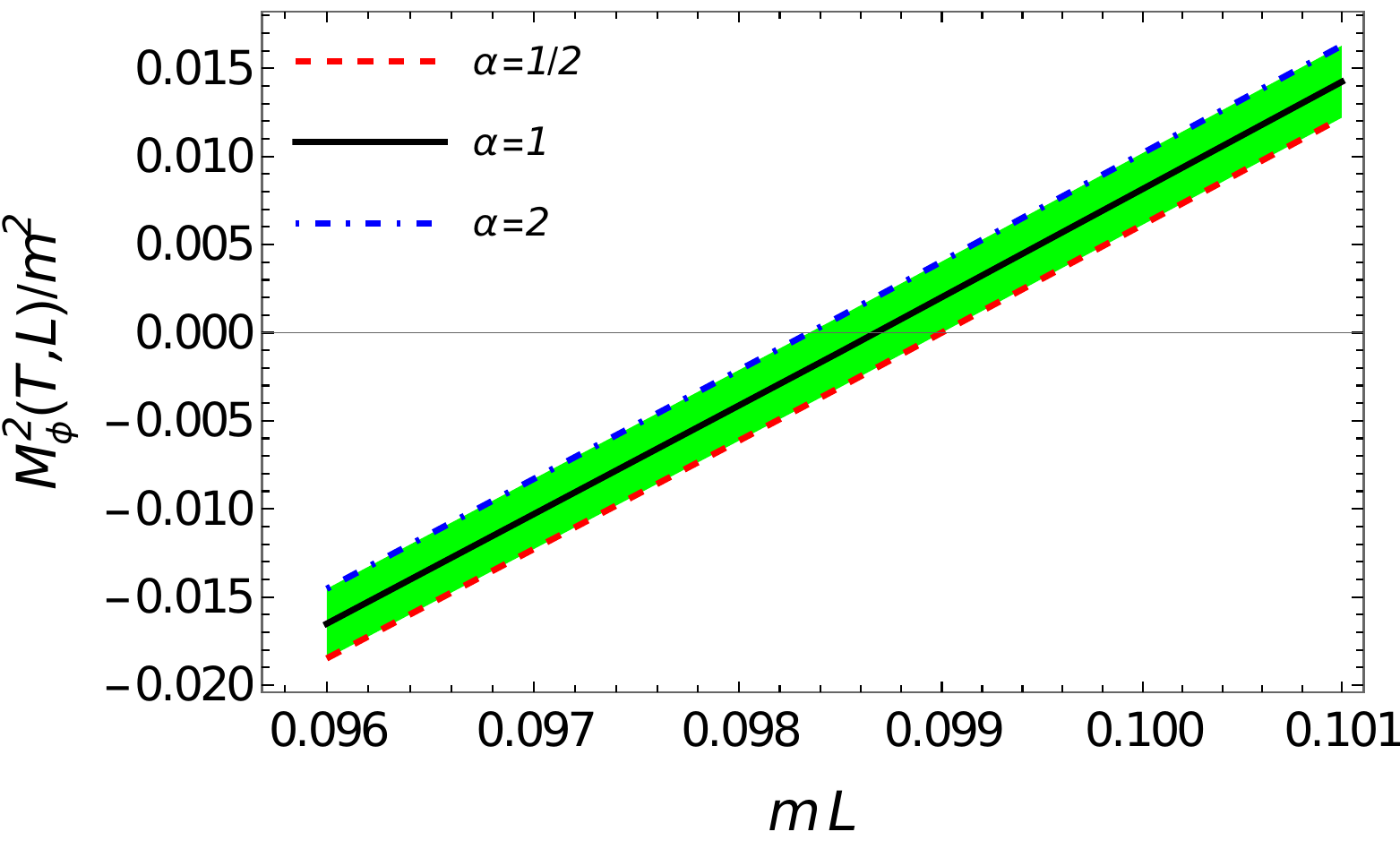}}
\subfigure[]{\includegraphics[width=5.7cm]{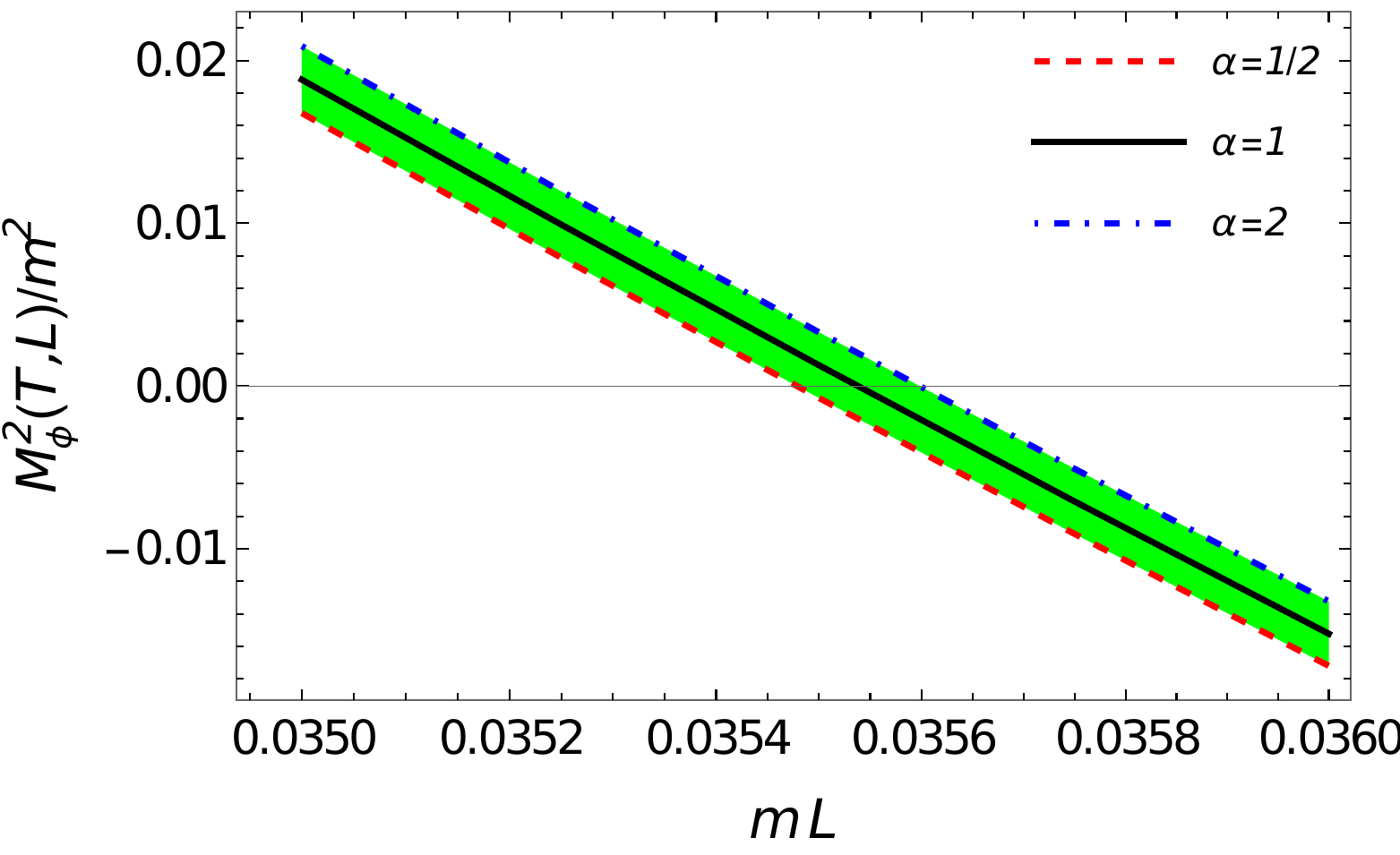}}
\subfigure[]{\includegraphics[width=5.7cm]{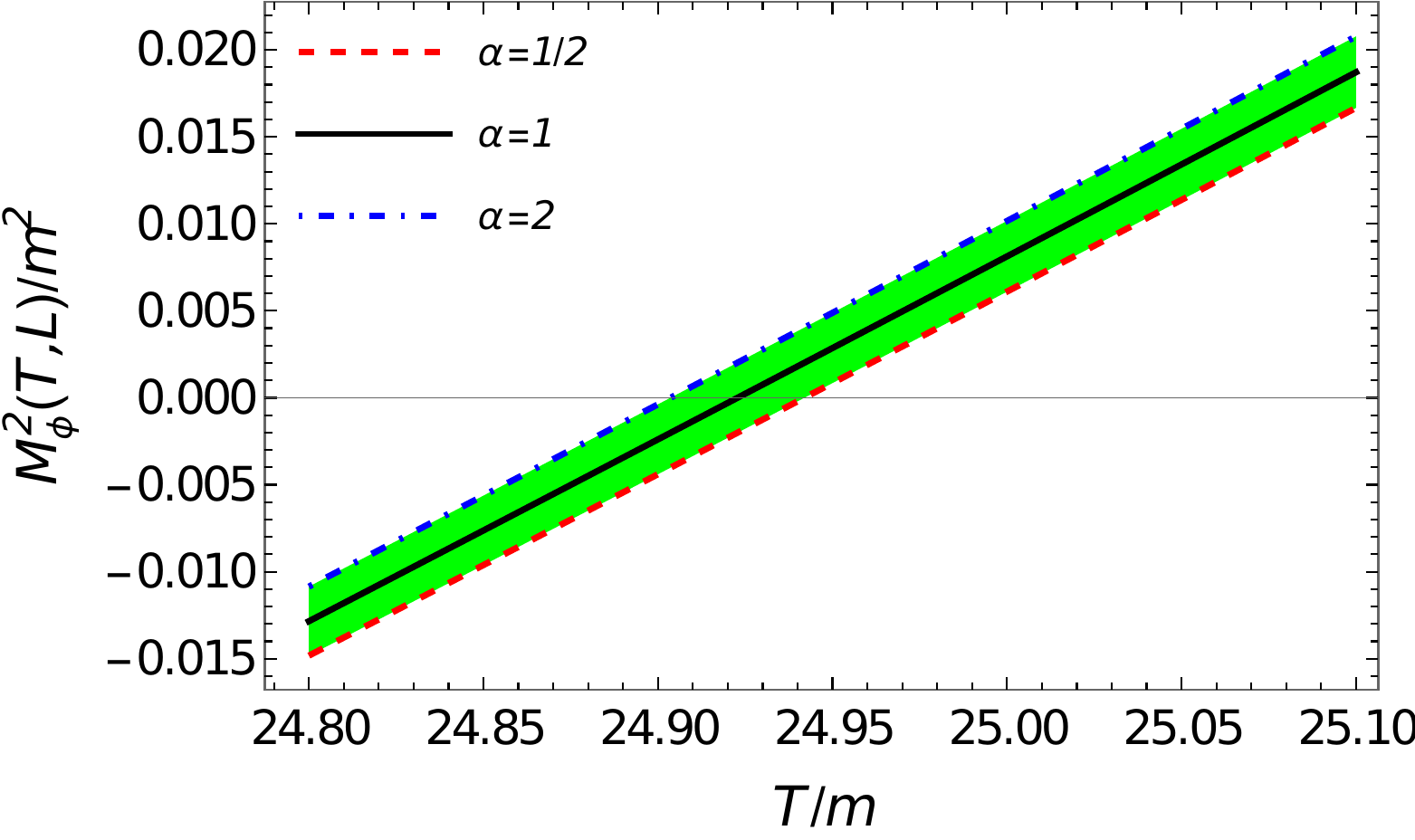}}
\caption{A detailed view of the variation of the critical points with respect
to changes in the renormalization scale around the reference
value $\mu_0$. The region around the two critical points in the length 
$L$ shown in  {}Fig.~\ref{fig2}(a) for $T/m(\mu_0) = 25$
are here shown in the panels (a) and (b). Panel (c) shows the region around 
the critical point in  the
temperature for the case with $m(\mu_0) L=0.1$ shown in {}Fig.~\ref{fig2}(b).
}
\label{fig3}
\end{figure*}
\end{center}

At this point, it becomes useful to analyze the dependence of any of
these results with the variation of the scale. Let us remainder that
a sensitive of the results with the renormalization scale $\mu$ can
also be seen as an indicator of the reliability of the approximation
method which is being used in our study, i.e., the bubble resummation
scheme explained in Sec.~\ref{secring}. {}For instance, the effective
mass here is directly related to the 1PI two-point function at zero
momentum, $\Gamma^{(2)}_i(0,0)\equiv - M_i^2$ and $\Gamma^{(2)}_i$
must satisfy the RG equation~\cite{Peskin:1995ev},
\begin{equation}
\mu \frac{d \Gamma^{(2)}_i}{d\mu} =0.
\label{RGGamma2}
\end{equation}
Of course, since we cannot evaluate $\Gamma^{(2)}_i$ to arbitrary precision,
the level of sensitive of it on the renormalization scale gives us an indication
of the level of accuracy of the method used in its derivation.

As already explained at the end of Sec.~\ref{scales}, we follow the standard
procedure usually adopted in many of the studies in the context of the hard thermal
loop and vary $\mu$ with respect to the reference scale by a factor of two,
$\mu/\mu_0 \equiv \alpha=[1/2,2]$. This typically gives a good indicator of how
relevant thermodynamical quantities (like the pressure) are sensitive to the 
scale~\cite{Andersen:2004fp}. Taking for instance the example shown in {}Fig.~\ref{fig2}
in the cases of $T/m(\mu_0) = 25$ and $m(\mu_0) L=0.1$ (solid line in 
 {}Figs.~\ref{fig2}(a) and ~\ref{fig2}(b), respectively), the effect of changing the
scale is very small. This is illustrated in {}Fig~\ref{fig3} by zooming in around each of the critical
points shown in there. Our results show that the considered variation of the renormalization
scale leads to a change in the critical points by around the one percent level and smaller. 
This is in a sense consistent with the fact that we are considering not large couplings
neither is the temperature is too large to cause a significant change here (recalling from
the flow equations that variations with the scale are logarithmic).

{}Figure~\ref{fig3} shows the behavior for the effective mass $M_\phi$
close to the transition points in the temperature and length.
The same can be done also for the field expectation value $\langle \phi \rangle$
shown in  {}Figs.~\ref{fig2}(c) and ~\ref{fig2}(d).
Let us recall that the inverse of the effective mass is equivalent
to the correlation length of the field, $\xi_\phi = 1/M_\phi$,
while $\langle \phi \rangle$ makes the role of the order parameter
for the phase transition, much the same way as the correlation
length and magnetization, respectively, for an Ising system in three spatial 
dimensions in the context of statistical physics~\cite{Parisi:1988nd}.
The way these quantities approach the critical point are specified by
the critical exponents $\nu$ and $\beta$, which can be defined as
\begin{eqnarray}
\nu = \lim_{\tau \to 0} \frac{\ln |M_\phi(\tau)|}{\ln|\tau|},
\label{nu}
\\
\beta = \lim_{\tau \to 0} \frac{\ln \langle \phi \rangle(\tau)}{\ln|\tau|},
\label{betac}
\end{eqnarray}
where, in the usual prescription, $\tau$ denotes the reduced temperature,
$\tau=(T-T_c)/T_c$ and $T_c$ is the critical temperature. In the present problem,
the critical point can be approached by varying either the temperature or the length $L$.
Hence, we can define $\tau$ with respect to either $T$ or $L$, according to the equivalence
between $T$ and $1/L$ already discussed at the end of Sec.~\ref{section3}, and make
estimates concerning the critical exponents $\nu$ and $\beta$. Already from 
{}Fig.~\ref{fig3} we can see that the effective mass square always approaches the critical
point in a linear way. Since we expect that $M_\phi^2 \propto |\tau|^{2 \nu}$ as $\tau\to 0$, 
hence, $\nu=1/2$. The same can be studied for how the expectation
value of the field approaches the critical points. Taking sufficiently small values for $\tau$,
we verify that here too we get $\beta=1/2$. These are the mean-field critical exponents and which
are expected for the level of approximation we are presently considering. Deviations
from the mean-field values are only expected to happen when the full two-loop order terms
are also resummed and to approach the expected critical exponents for this problem can only
be achieved through a three-loop and higher-order resummation scheme (see, e.g., 
Ref.~\cite{Kleinert:2001hn} for some examples). 

\begin{center}
\begin{figure}[!htb]
\includegraphics[width=7.5cm]{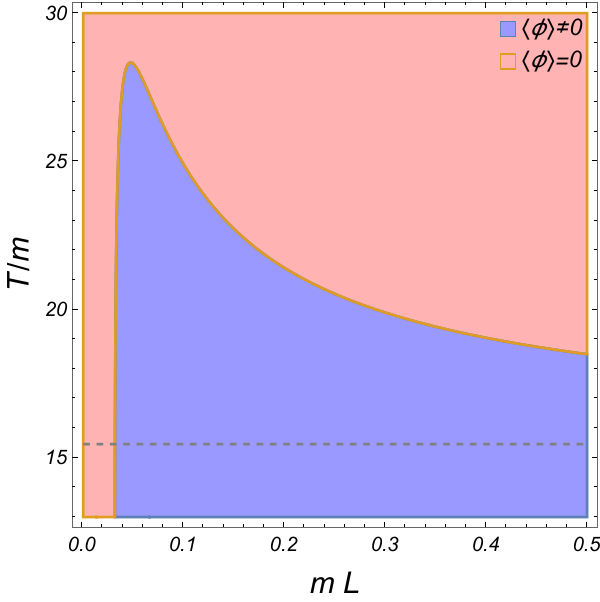}
\caption{The phase diagram for the one-field case in the plane $(L,T)$
  for the case of the same parameters used in {}Fig.~\ref{fig2}. The
  horizontal dashed line indicates the critical temperature in the
  case of $L\to \infty$, $T_c(L\to \infty)/m\simeq 15.5$.}
  \label{fig4}
\end{figure}
\end{center}

{}Finally, for illustration, the
different transition behaviors displayed in {}Fig.~\ref{fig2} are
shown in the phase diagram depicted  in {}Fig.~\ref{fig4}, where the
regions of symmetry breaking, $\langle \phi \rangle \neq 0$, and
symmetry restored, $\langle \phi \rangle = 0$, are shown in the plane
$(L,T)$.

The variation of the renormalization scale in the same range considered
in {}Fig.~\ref{fig3} only leads to very small changes (at most at the one percent level) 
in the critical curve shown in {}Fig.~\ref{fig4} and, therefore, it is not shown
explicitly.
The phase diagram depicted in {}Fig.~\ref{fig4} shows that the
critical temperature fast increases when $L$ is increased when
starting from very small values ($mL \gtrsim 0.03$), reaches a
maximum at $mL\simeq 0.05$ and then decreases towards large $L$. The
horizontal dashed line indicates the critical temperature in the case
of the absence of finite-size effects ($L\to \infty$), whose value for
the parameters used in {}Fig.~\ref{fig4} is found to be $T_c(L\to
\infty)/m \simeq 15.5$.  The behavior analogous to 
reentrant phase transitions as a function
of $L$ are clearly visible when taking constant temperature values and
that happens with temperatures in the range  $15.5 \lesssim T/m
\lesssim 28$.

The phase structure behavior displayed in {}Fig.~\ref{fig4} is easy to understand when we
look at the temperature and size dependence of the loop integral $I_2^{(1)}$ discussed at the
end of Sec.~\ref{section3}. The reentrant behavior is directly related to the
fact that the loop integral $I_2^{(1)}$ displays a minimum for ranges of values for
$T$ and $L$ as shown in {}Fig.~\ref{fig1}. {}For fixed $L$ and increasing $T$, as
$I_2^{(1)} \propto T^2$, eventually SR is realized. {}Finally, when 
$L$ is decreased with fixed $T$, again we see that since now $I_2^{(1)} \propto 1/L^2$,
then the symmetry will also be restored for sufficiently small values of $L$, as it is
explicitly seen in {}Fig.~\ref{fig4}.

\subsection{The two-scalar field case}

Let us now study the multiscalar field case, i.e., including both
$\phi$ and $\chi$ with a nonvanishing intercoupling $\lambda$ between
them.  We are mainly interested in exploring a parameter space where
the usual SR can happen at high temperatures
along a given field direction, while the other field experiences ISB.
Let us first recall that from the PT estimates from 
Sec.~\ref{section2}, considering a negative intercoupling $\lambda$,
when Eq.~(\ref{isbcondition}) is observed, but still satisfying the 
boundness condition Eq.~(\ref{boundness}), we expected an ISB
phase transition in the {\it i}-field direction, when $m_i^2 >0$,
or, equivalently, a SNR, when $m_i^2<0$. According to the previous
discussions, these phase transition patterns are expected at either
at high temperatures (at fixed $L$), or, likewise, at small values
of the length, $L \ll 1/m,1/T$ (at fixed $T$). By focus on the ISB case,
let us investigate first the parameter space that leads to this
phase transition pattern. 

\begin{center}
\begin{figure}[!htb]
\includegraphics[width=7.5cm]{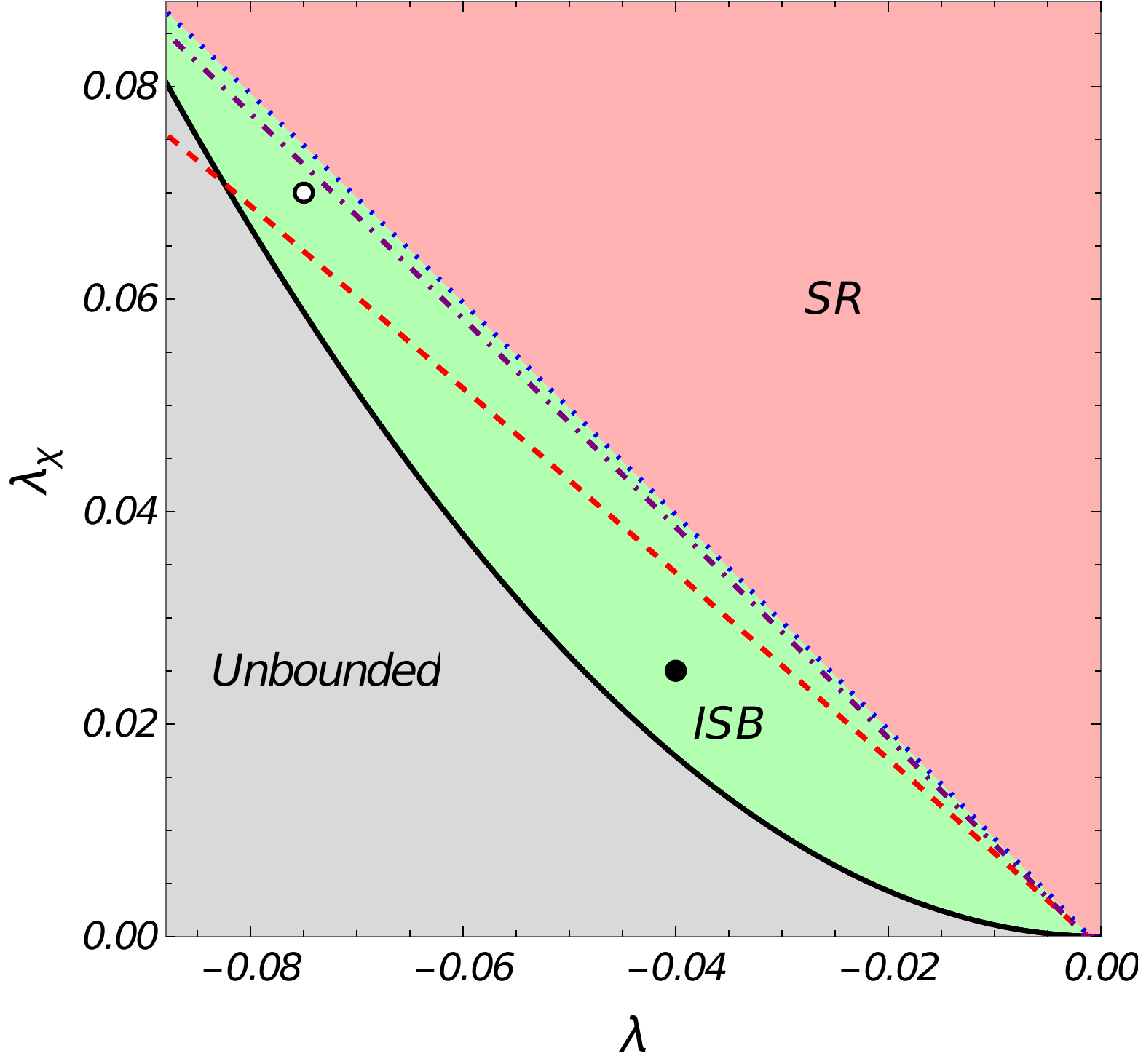}
\caption{ The parameter space able to lead to inverse symmetry breaking
in the case of a fixed value of $\lambda_\phi=0.8$. 
The black and white (open) dots in the figure indicate, respectively, the locations for the parameter
sets I and II used in our numerical studies. The solid line delimits the region
satisfying the boundness of the potential. The dotted and dashed lined delimits
the separation between the SR and ISB phases in the cases of PT and the nonperturbative
bubble resummation procedure in the case of the bulk ($L\to \infty$) and for 
a sufficiently high temperature. The dash-dotted line in between them indicates
a boundary set when using a finite value for $L$, $m L=0.075$. }
  \label{fig5}
\end{figure}
\end{center}

Without loss of generality, we can consider, for
example, the two sets of renormalized couplings\footnote{Unless stated otherwise,
we work with all parameters defined at the reference scale $\mu_0$.
We could likewise work directly with the scale independent parameters,
the barred quantities of Sec.~\ref{section3}. {}For the coupling 
constants that we will be considering the difference between the barred
parameters and unbarred ones (which explicitly depend on the scale)
are always very small and the use of one or another set of parameters
does not affect our results in any qualitative or significant quantitative
way.}: 
Set I: $\lambda_\phi=0.8,\, \lambda_\chi=0.025$ and
$\lambda=-0.04$; and
Set II: $\lambda_\phi=0.8,\, \lambda_\chi=0.07$ and
$\lambda=-0.075$ , along also with the initial choice $N_\phi=N_\chi=1$.
Both sets of couplings satisfy Eq.~(\ref{boundness}).  
Then, with the perturbation equations in the high-temperature approximation, 
e.g., Eq.~(\ref{MiT}),  they predict an ISB
phase transition along the direction of the $\chi$ field whenever
$m_\chi^2 >0$. On the other hand, for $m_\chi^2 <0$, SNR should
manifest, in which case the $\chi$-field expectation value should
always remain as nonvanishing, $\langle \chi \rangle \neq 0$. The
symmetry behavior along the $\phi$ field direction depends only on the
sign of its mass square term entering in the Lagrangian density. {}For
$m_\phi^2 >0$, it (in the $\phi$ field direction) remains in a
symmetry restored phase,  $\langle \phi \rangle = 0$, while
considering initially (at $T=0$) a symmetry broken (SB) phase along
the $\phi$ field direction, i.e., $m_\phi^2 <0$, the usual symmetry
restoration at high temperatures is expected.  Of course, had we
chosen different assignments for the couplings, the roles of the
$\phi$ and $\chi$ fields are expected to be reversed under the
phenomena of SB/SR and ISB/SNR.  {}For definiteness, in the analysis
that follow, we will consider that $\phi$ suffers the usual SR at high
temperatures, while $\chi$ should experience ISB, i.e., we choose the
mass renormalized parameters such that  $m_\phi^2 <0$ and  $m_\chi^2 >0$ and
under the above given choice of representative values for the two sets of coupling
constants.  Of course, other combinations of parameters could be used
but the results could be similarly interpreted.

It is useful to show the two sets of couplings that we are considering here
in a given plane in the coupling constants parameter space. With $\lambda_\phi=0.8$.
in {}Fig.~\ref{fig5}, we present the parameter available for ISB. The location 
of the sets I and II are represented by the black dot and white (open) points, respectively. 
The region in between the solid and dotted lines is the region for ISB predicted by
PT according to Eq.~(\ref{isbcondition}) and when $L\to \infty$.
The region for ISB according to the nonperturbative results coming from the solution of
the gap equations is, however, delimited by the solid and dashed lines (also when
$L\to \infty$). Note that there is reduction of the parameter region available
for ISB when compared to PT\footnote{This figure can be compared for
instance to the similar one of Ref.~\cite{Pietroni:1996zj} (given by {}Fig.~1 in that reference), 
which considered
instead the method of the exact renormalization group to analyze ISB. The reduction
of the parameter region is equivalent to the one we obtain here.}. The dash-dotted line
delimits the region for ISB for the case of finite $L$, which is taken as $L= 0.075/m$
in the example shown in {}Fig.~\ref{fig5}. Note that the finite-size effect is to enlarge
the region for ISB when compared to the result at $L\to \infty$.

\begin{center}
\begin{figure}[!htb]
\subfigure[Parameter set I]{\includegraphics[width=7.5cm]{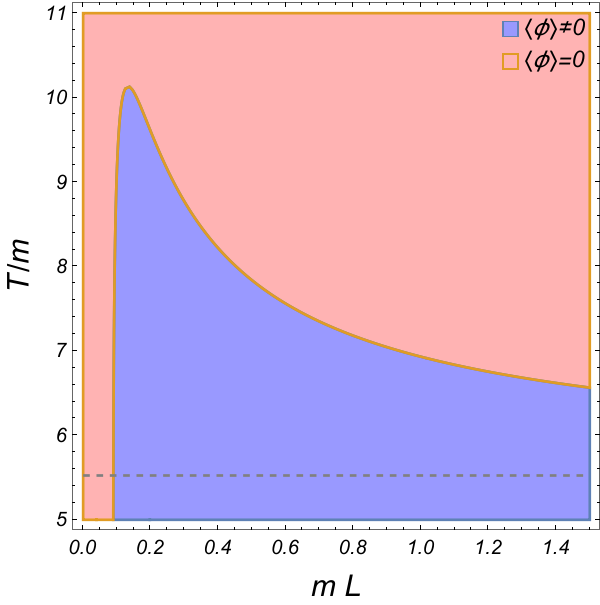}}
\subfigure[Parameter set II]{\includegraphics[width=7.5cm]{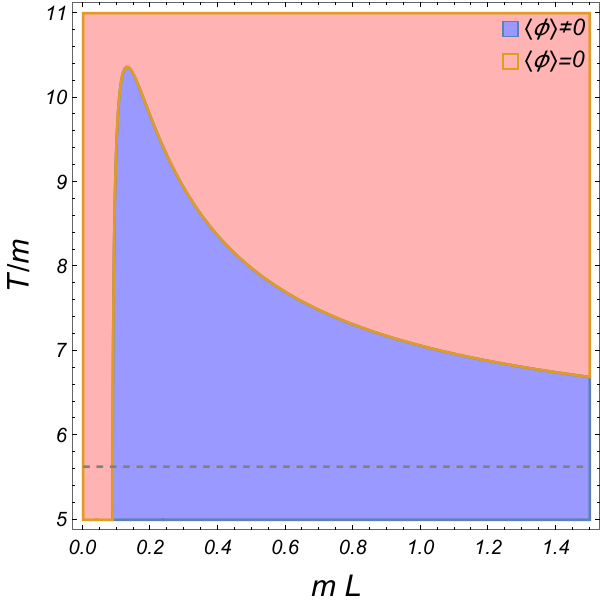}}
\caption{The phase diagram for $\phi$ in the coupled field case in the
  plane $(L,T)$. The horizontal dashed line indicates the critical
  temperature in the case of $L\to \infty$, $T_c(L\to \infty)/m\simeq
  5.52$ (set I) and $T_c(L\to \infty)/m\simeq
  5.63$ (set II). }
  \label{fig6}
\end{figure}
\end{center}

In both parameter sets I and II we expect that the field $\phi$ will experience the usual SR as
the temperature increases. Also, as in the one-field case studied in the previous subsection,
as we decrease $L$, SR is also expected. This is explicitly illustrated in {}Fig.~\ref{fig6},
where the transition pattern for $\phi$ in both sets of parameters is shown.
{}Figure~\ref{fig6} shows again the characteristic reentrant (double transition point)
like behavior as the compactification size $L$ is changed and a
maximum critical temperature that can be reached when $L$ changes. 
Note that in the results shown in {}Fig.~\ref{fig6} and the next ones, we have
verified explicitly that the change in the
scale $\mu$ only leads to small corrections, just like in the one-field case, and causes no
change in our conclusions regarding the phase transition patterns reported here. Thus, 
all the following results are presented only at the reference scale $\mu_0$.

\begin{center}
\begin{figure}[!htb]
\subfigure[Parameter set I]{\includegraphics[width=7.5cm]{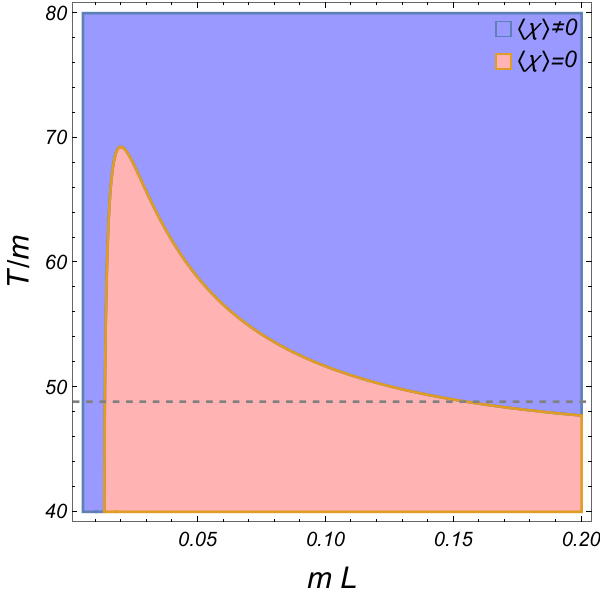}}
\subfigure[Parameter set II]{\includegraphics[width=7.5cm]{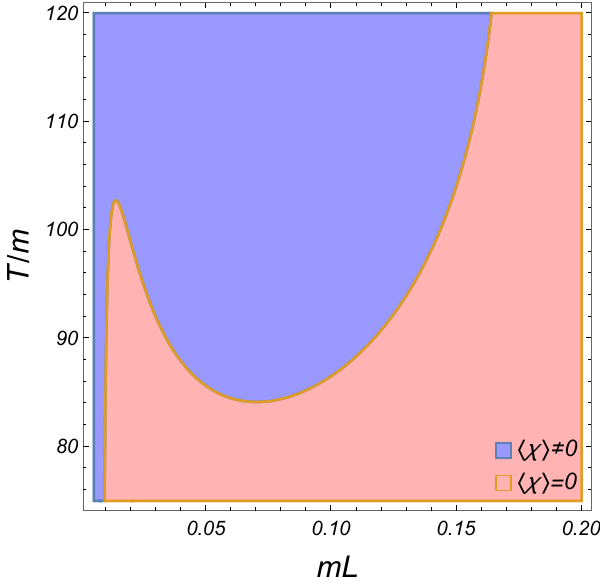}}
\caption{The phase diagram for $\chi$ in the coupled field case in the
  plane $(L,T)$. The horizontal dashed line in the case of panel (a) indicates the critical
  temperature in the case of $L\to \infty$, $T_c(L\to \infty)/m\simeq
  48.8$.}
  \label{fig7}
\end{figure}
\end{center}

The situation becomes more interesting when we now analyze the transition pattern for
the field $\psi$. In the parameter set I we expect the field $\psi$ to display ISB
just like in the case of the bulk case ($L\to \infty$). This is indeed the case as
shown in {}Fig.~\ref{fig7}(a) when $L$ is increased. However, as expected, ISB
also manifests in the small $L$ region, for the same reason as discussed for the transition
in the $\phi$ direction (though now instead of SR, we have ISB) and
it is again due to the behavior displayed by the loop function in terms of $L$ and $T$.
Note that now, we have a reentrant transition to a symmetry restored phase in between the
two extremes.  However, for the parameter set II, which resides in between the dashed line
($L \to \infty$) and the finite $L$ region shown in {}Fig.~\ref{fig5}, we do not expect
ISB to persist for large values of $L$, as the dash-dotted line moves towards the dashed line
and the point will reside in the SR phase. Thus, here, ISB is truly only a consequence
of the finite-size effect as illustrated in {}Fig.~\ref{fig7}(b).

\begin{center}
\begin{figure}[!htb]
\subfigure[]{\includegraphics[width=7.5cm]{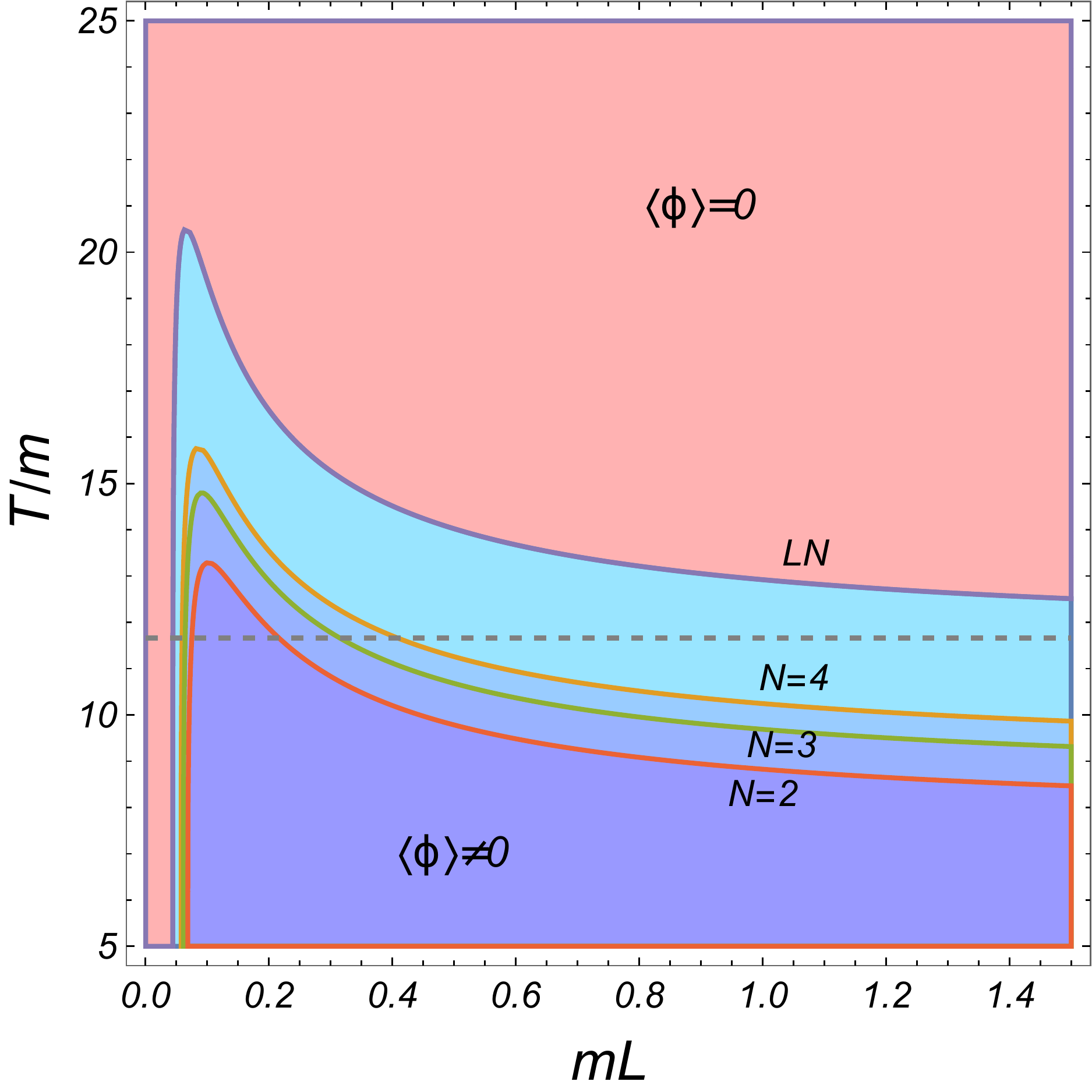}}
\subfigure[]{\includegraphics[width=7.5cm]{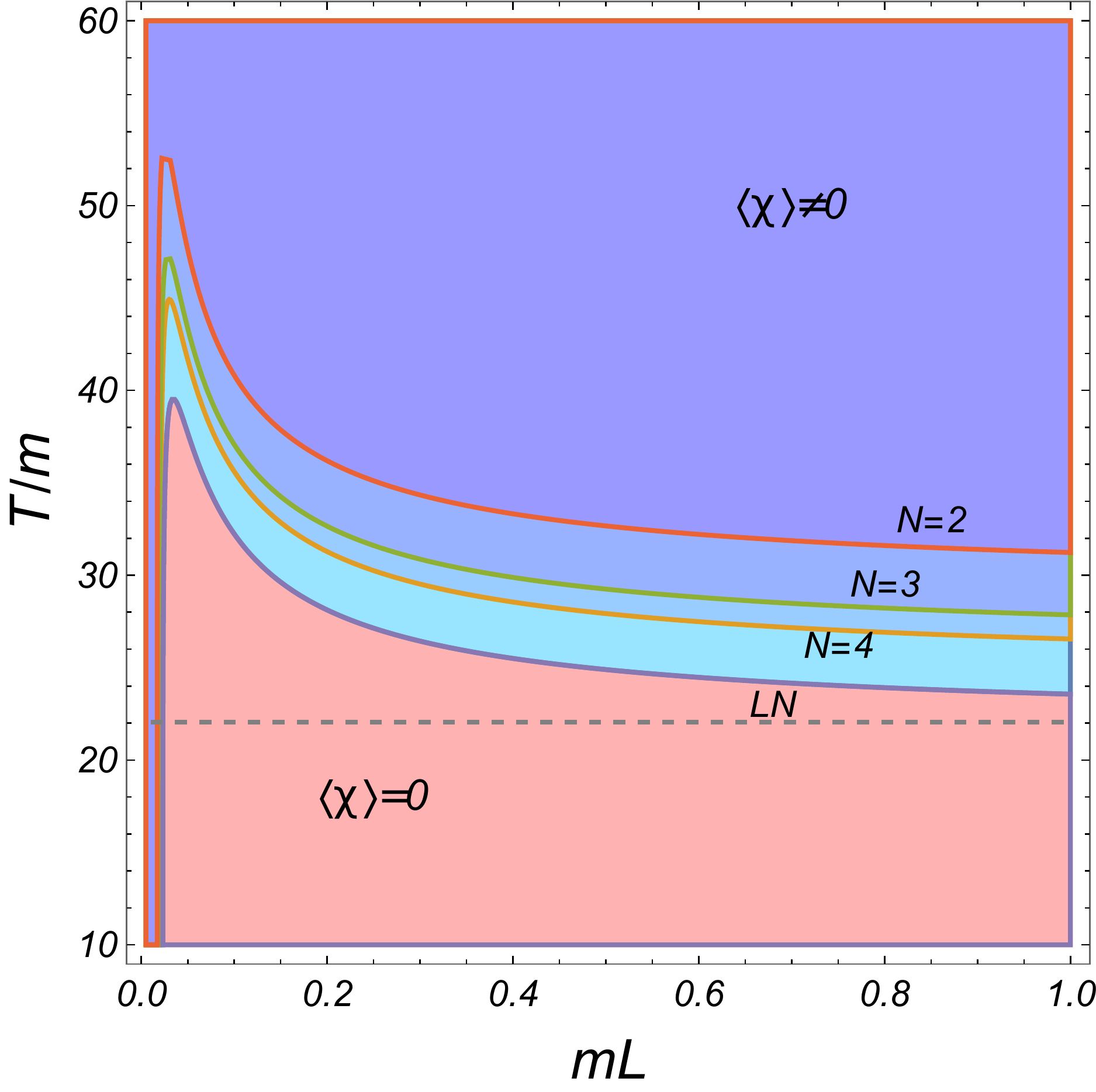}}
\caption{The phase diagram for $\phi$ (panel a) and for $\chi$ (panel
  b) in the coupled field case in the plane $(L,T)$ using the LN
  approach. The parameters chosen are the same as the set II considered in
  {}Figs.~\ref{fig6} and \ref{fig7},  $m_\phi^2=-m_\chi^2=m^2>0$,
  $\lambda_\phi=0.8,\, \lambda_\chi=0.07$ and $\lambda=-0.075$.  The
  horizontal dashed line indicates the critical temperature in the LN
  approximation and in the bulk case ($L\to \infty$), $T_{c,\phi}^{\rm
    LN}(L\to \infty)/m\simeq 11.7$ and $T_{c,\chi}^{\rm LN}(L\to
  \infty)/m\simeq 22$. }
  \label{fig8}
\end{figure}
\end{center}

\section{Finite $N$ effects and comparison with the large-$N$ approximation}
\label{largeN}

So far, in the previous section we have restricted to cases where
$N_\phi=N_\chi=1$, i.e., in the context of a two-field model with
symmetry $Z_2 \times Z_2$. Let us now investigate how those results
might get affected  by finite $N$ effects. It is useful in this
context to compare the finite $N$ results with the ones derived in the
context of the large-$N$ approximation.

In the large-$N$ (LN) approximation  (for a review, see, e.g.,
Ref.~\cite{Moshe:2003xn}) both coupling constants and fields are
normalized by $N$, such that, for instance, $\lambda_i \to
\lambda_i/N_i$, $\lambda \to \lambda/\sqrt{N_\phi N_\chi}$, $\phi\to
\sqrt{N_\phi}\phi$ and $\chi \to \sqrt{N_\chi}\chi$, while the
rescaled coupling are kept fixed as $N_i \to \infty$.  Without loss of
generality, we will consider $N_\phi=N_\chi \equiv N$. Note that in
the LN approximation, the simple one-loop expressions, e.g.,
Eqs.~(\ref{MphiTL}) and (\ref{MchiTL}), become exact, since higher
order terms become suppressed by factors of $1/N_i$.  Next, we will
compare the LN approximation result for the phase diagram in the plane
$(L,T)$ for $\phi$ and $\chi$ with the corresponding finite $N$ ones
constructed in this approach. In {}Fig.~\ref{fig8}(a) we show the
result for the phase diagram for $\phi$, while in
{}Fig.~\ref{fig8}(b) it is shown the result for $\chi$. We have used
the values $N=2$, $N=3$ and $N=4$ for comparison, which can be
motivated by, e.g., coupled complex scalar fields with $O(2)\times
O(2)$ symmetry, coupled Heisenberg type of models  with $O(3)\times
O(3)$ symmetry, or a coupled linear $\sigma$-model  with $O(4)\times
O(4)$ symmetry, respectively.  The set of parameters considered was 
set II, which was used in previous subsection. We note that for the 
large-$N$ case, as also for the finite $N$ examples shown in {}Fig.~\ref{fig8},
the point represented by the parameter set II moves to the ISB inner
region shown in {}Fig.~\ref{fig5}. Hence,  as oppositely seen
in {}Fig.~\ref{fig7} in the case of $N_\phi=N_\chi=1$, here ISB persists 
even at larger values of $L$. We also note from the results shown in 
{}Figs.~\ref{fig8} that in the LN approximation
we still observe reentrant phases with double critical values for $L$
in both the $\phi$ and $\chi$ directions.

\section{Conclusions}
\label{conclusions}

In this work we have investigated the possible phase transitions
patterns in a two-field multicomponent  scalar model with symmetry
$O(N_\phi)\times O(N_\chi)$ when both thermal and finite-size effects
are present. While in the bulk (in the absence of space boundaries)
phenomena like ISB have shown  to be present at high temperatures,
when the intercoupling between the fields is negative, the  phase
transition patterns when finite-size effects are present become more
involved.  The finite-size effects allow the system to display
behavior analogous to 
reentrant symmetry breaking (which can happen in between two
phases of symmetry restoration) and vice versa, where symmetry
restoration can happen in between two phases of symmetry breaking.
These type of phenomena can happen for both fields depending on the
choice of parameters and they are a consequence of the behavior
of the one-loop integral as a function of $T$ and $L$, which presents
a minimum depending on the choice of temperature and length.  
In this work we have analyzed these phase transition phenomena
with the use of the bubble (or ring) resummed gap equations for both
the effective masses for the fields and for their expectation values,
$\langle \phi\rangle$ and $\langle \chi\rangle$. The results were also
compared with the large-$N$ approximation. 
The  behavior analogous to 
reentrant symmetry breaking for the fields was shown to be present in both
cases. 

We have considered in this paper the Dirichlet boundary condition for
the space compactification in a slab geometry in the three-dimensional
space, $\mathbb{R}^2 \times [0,L]$.  The Dirichlet boundary condition
is a well-motivated condition for different physical systems as has
been argued in the recent literature. {}For practical purposes, it is
also computationally convenient since there are no zero modes when
working with the discrete frequencies and, consequently, direct
simple equations in the limit of large temperature can be derived. We
have obtained the expressions for the two-point one-loop self-energy
correction contributing to the effective masses and we have also made
use of the renormalized parameters (masses and coupling constants). 
We have shown that typical variations of the scale around a reference
value do not change in any significant way the transition patterns
describe and found in this work, 
with the variations with the scale remaining small. This fact can be
used as an additional argument concerning the reliability of the approximation used.
Our results can, in principle, also be extended to higher loops, which 
would allow eventually to explore regions of larger couplings than the ones
we have considered here.

Other types of geometries can also be considered as future
developments of this work. But still, the  simpler slab geometry here
considered can be of interest in practical physical applications, most
notably, like in condensed matter problems where, for example, the
thickness of thin films are
important~\cite{material1,material2,material3,material4,material5}.  With the
advent of even more  miniaturized electronic devices, it is extremely
relevant to analyze how the size and interface effects change the
properties and performances of nanomaterials.  In particular, similar
phase transition behaviors as a function of the thickness  that
resembles the ones we have seen here, have also been observed in
different materials. {}For instance, a rapid grow of the critical
temperature with the thickness and then a smooth decrease of $T_c$, as
it appears in {}Figs.~\ref{fig2}, \ref{fig6} and \ref{fig8}, for
example, have been experimentally measured in different materials
displaying superconductivity transitions (see,
e.g.,~Ref.~\cite{material6}). In this context, even though we have
here considered a relativistic type of model, our results can still be
of interest in applications in the condensed matter context. {}For
instance, excitons type of systems exhibit a relativistic dispersion
relation~\cite{Lopes:2019mjc} and, furthermore, are effectively
modeled as a multifield scalar model for which our results can be
directly applied to. 
Our results can also be of interest for high-energy relativistic systems. 
Relativistic type of systems typically
modeled by multifield scalar models with both inter and intracouplings
include for example those in the context of
color-flavor superconductivity~\cite{Alford:2007qa,Andersen:2008tn,Silva:2023jrk},
which the study of the effect of space compactifications can be of
relevance.  
In addition, other applications are also possible, like, for example,
to infer possible finite-size effects in the transition from a quark-gluon
plasma to hadronic matter in the case of the fireball formed in the collision of
heavy ions. The pancake like shape of the formed plasma (for a review, see, e.g.. Ref.~\cite{Nguyen:2020zux}) 
suggest that at the
central region our slablike geometry might be relevant and these consequent
finite-size effects to be important in the interpretation of the transition
and thermodynamics of the plasma, as suggested, e.g., in Ref.~\cite{Mogliacci:2018oea,Abreu:2019czp,Abreu:2021btt}.
In addition, other system of possible relevance could be in the physics
of compact stars. The structure of hybrid type of stars typically involves
crust like structures which is important in the description of the
stability and structure of these type of compact stellar objects (see, e.g., Ref.~\cite{Lopes:2022efy}
and references therein).
It is conceivable that in the presence of more than one crust type of
structure, if thin enough, finite-size effects as the ones studied here 
can also be important.
{}Finally, while we have investigate the transition patterns in a thermal environment, the same
can also be performed in the context of quantum phase transitions, which also find applications
in the context of multifield type of models like the one studied here~\cite{Heymans:2022idr}.
The study of finite sizes in these type of systems can also be of interest as well.

Applications in the context of any of the above mentioned type of systems and others require, of course,
a separate dedicated study. Other extensions of our work can also
include the uses of other boundary conditions and the analysis under
different nonperturbative methods. These further developments are clearly of interest and we hope
to address some of them in the future.

\begin{acknowledgements}

R.O.R. would like to thank the hospitality of the Department of Physics
McGill University.
The authors acknowledge financial support of the Coordena\c{c}\~ao de
Aperfei\c{c}oamento de Pessoal de N\'{\i}vel Superior (CAPES) -
{}Finance Code 001. R.O.R. is also partially supported by research
grants from Conselho Nacional de Desenvolvimento Cient\'{\i}fico e
Tecnol\'ogico (CNPq), Grant No. 307286/2021-5, and from
{}Funda\c{c}\~ao Carlos Chagas Filho de Amparo \`a Pesquisa do Estado
do Rio de Janeiro (FAPERJ), Grant No. E-26/201.150/2021. 

\end{acknowledgements}


\end{document}